\documentclass[12pt, onecolumn, draftcls]{IEEEtran}
\usepackage{amsmath,amssymb,color,amsthm,cite,float}
\usepackage{amsfonts, mathtools, bm, comment, cite, tikz, graphics, graphicx}
\usetikzlibrary{decorations.pathreplacing,calligraphy}
\usepackage{}
\usepackage{caption}
\usepackage{subcaption}
\usepackage{array}
\usepackage{multirow}
\usepackage{wasysym}
\usepackage{enumerate}
\usepackage{makecell}
\usepackage{tabularx} % in the preamble
\usepackage{algorithm,algcompatible}
\algnewcommand\INPUT{\item[\textbf{Input:}]}%
\algnewcommand\OUTPUT{\item[\textbf{Output:}]}%
\newcommand\numberthis{\addtocounter{equation}{1}\tag{\theequation}}
\usepackage[mathscr]{eucal}
\DeclareMathAlphabet{\mathdutchcal}{U}{dutchcal}{m}{n}
\SetMathAlphabet{\mathdutchcal}{bold}{U}{dutchcal}{b}{n}
\DeclareMathAlphabet{\mathdutchbcal}{U}{dutchcal}{b}{n}

 \usepackage{setspace}
\begin{document}

\title{Block Orthogonal Sparse Superposition Codes for Ultra-Reliable Low-Latency Communications}

\author{Donghwa~Han,~\IEEEmembership{Graduate~Student~Member,~IEEE,}
       Jeonghun~Park,~\IEEEmembership{Member,~IEEE,} Youngjoo Lee,~\IEEEmembership{Member,~IEEE,} 
       H. Vincent Poor,~\IEEEmembership{Fellow,~IEEE,} 
        and~Namyoon~Lee,~\IEEEmembership{Senior~Member,~IEEE}
	 
% \author{Donghwa~Han,~\IEEEmembership{Graduate~Student~Member,~IEEE,}
%         Bowhyung~Lee,~\IEEEmembership{Graduate~Student~Member,~IEEE,}
%         Geon~Choi,~\IEEEmembership{Graduate~Student~Member,~IEEE,}
%         and~Namyoon~Lee,~\IEEEmembership{Senior~Member,~IEEE}% <-this % stops a space
\thanks{D. Han and Y. Lee are with the Department of Electrical Engineering, Pohang University of Science and Technology (POSTECH), Pohang 37673, South Korea (e-mail: \{dhan92, youngjoo.lee\}@postech.ac.kr).}% <-this % stops a space
\thanks{J. Park is with the School of Electronics Engineering, College of IT Engineering, Kyungpook National University, Daegu 41566, South Korea (e-mail: jeonghun.park@knu.ac.kr).}
\thanks{H. V. Poor is with the Department of Electrical and Computer Engineering, Princeton University, NJ, USA (e-mail: poor@princeton.edu).}
\thanks{N. Lee is with the School of Electrical Engineering, Korea University, Seoul 02841, South Korea (e-mail: namyoon@korea.ac.kr).}

\thanks{This paper was presented in part at IEEE GLOBECOM 2021 \cite{BOSS1}.}}
\maketitle

\setlength\arraycolsep{2pt}
\makeatletter 
\makeatother
\vspace{-12mm}

\begin{abstract}
Low-rate and short-packet transmissions are important for ultra-reliable low-latency communications (URLLC). In this paper, we put forth a new family of sparse superposition codes for URLLC, called block orthogonal sparse superposition (BOSS) codes. We first present a code construction method for the efficient encoding of BOSS codes. The key idea is to construct codewords by the superposition of the orthogonal columns of a dictionary matrix with a sequential bit mapping strategy. We also propose an approximate maximum a posteriori probability (MAP) decoder with two stages. The approximate MAP decoder reduces the decoding latency significantly via a parallel decoding structure while maintaining a comparable decoding complexity to the successive cancellation list (SCL) decoder of polar codes. Furthermore, to gauge the code performance in the finite-blocklength regime, we derive an exact analytical expression for block-error rates (BLERs) for single-layered BOSS codes in terms of relevant code parameters. Lastly, we present a cyclic redundancy check aided-BOSS (CA-BOSS) code with simple list decoding to boost the code performance. Our experiments verify that CA-BOSS with the simple list decoder outperforms CA-polar codes with SCL decoding in the low-rate and finite-blocklength regimes while achieving the finite-blocklength capacity upper bound within one dB of signal-to-noise ratio.

%Short-packet transmission is a key enabler for the upcoming wireless communications systems, notably ultra-reliable low-latency communications (URLLC) use cases in beyond 5G. In this paper, we propose block orthogonal sparse superposition (BOSS) codes for efficient short-packet communications. BOSS codes are a class of sparse superposition codes, and information bits are mapped into non-zero index locations of sparse message vectors. In BOSS codes, codewords are constructed as a superposition of mutually orthogonal sub-codewords. Harnessing this inherent orthogonality, a noise-tolerant decoding method is proposed for the additive white Gaussian noise channel. Decoding information communicated in BOSS codewords is equivalent to finding non-zero coefficients of a message vector, and the proposed decoder performs element-wise maximum a posteriori (MAP) decoding to identify the non-zero support. We analytically derive upper bounds on the block error probability of few-layered BOSS codes as a function of blocklength and code rate. In addition, it turns out that a single-block single-layer BOSS code achieves the ultimate Shannon limit in the power-limited regime, even with linear decoding complexity. Numerical results are provided to show that BOSS codes under the proposed MAP-based decoder outperform commercially-used channel codes in the low-rate short-blocklength regime, attesting to its feasibility and applicability to various URLLC applications.

%\begin{IEEEkeywords}
%Sparse superposition codes, short-packet transmissions, low-rate codes.
%\end{IEEEkeywords}
\end{abstract}

\section{Introduction}

\subsection{Motivation}

The aspiration for ultra-reliable low-latency communications (URLLC) is unrelenting for next generation wireless systems. URLLC is envisioned to ensure that a data packet is delivered within a very short time duration (e.g., within 1 ms) while satisfying very high reliability (e.g., the packet error rate is less than $10^{-6}$) \cite{URLLC_concept1_Popovski2014, URLLC_concept2_tactileInternet, URLLC_concept3, URLLC_concept4_Popovski2018, URLLC_concept5}. These stringent requirements of the latency and reliability are indispensable to support various mission-critical applications that demand prompt responses with ultra-high accuracy, including automated driving, industrial automation, and telesurgery \cite{URLLC_industrialComm., URLLC_IoT}. To deliver a data packet with extremely low-latency and high-reliability, it is essential to design a novel low-rate coded modulation technique that is fast decodable while achieving near-optimal performance in the finite-blocklength regime \cite{URLLC_channelCoding1, URLLC_channelCoding2}. The standard approach to designing low-rate codes at a short-blocklength is to concatenate standard codes  (e.g., turbo \cite{turbo} and low-density parity-check (LDPC) codes \cite{LDPC_Gallagar, LDPC_MacKay}) with a simple repetition code. For instance, the narrow-band internet-of-things standard allows up to 2048 repetitions of a turbo code with rate $1/3$ to meet the maximum coverage requirement \cite{NB-IoT1}. This simple construction, however, is very far from optimal when the blocklength is short.

Significant progress has been made on finite-blocklength information theory, which characterizes achievability and converse bounds for the highest channel code rate achievable at a given blocklength and error probability \cite{Polyanskiy_finiteCapacity}. Notwithstanding, designing the optimal code in finite-blocklength is very challenging because of the non-asymptotic behavior of codes in a finite-dimensional space. Recently,  Ar{\i}kan, the inventor of polar codes \cite{polarCode_SC}, presented a novel family of polar codes called \textit{polarization-adjusted convolutional (PAC) codes} \cite{PAC_Fano}. Unlike the polar code serially concatenated with a cyclic redundancy check (CRC) code used in 5G NR \cite{polarCode_List}, PAC codes take the convolutional transform with proper rate-profiling. PAC codes with a tree-search based sequential decoder were shown to achieve the finite-length converse bound very tightly, i.e., the Gaussian dispersion bound \cite{Polyanskiy_finiteCapacity}. However, these tree-search sequential decoders (e.g., Fano or stack decoders) are implemented with prohibitively high computational complexity at low signal-to-noise ratio (SNR). More importantly, decoding latency is unpredictable by the inherent nature of the tree-search based sequential decoding, which are not suitable for low-latency communication applications. 

In this paper, we take a different direction toward designing low-rate codes in finite-blocklength. Harnessing the power of sparsity in low-rate code design, we introduce a new class of low-rate codes for URLLC called \textit{block orthogonal sparse superposition (BOSS) codes}. Through the paper, we show that our BOSS code can achieve the Gaussian dispersion bound within one dB in the short-blocklength regime with a fast low-complexity decoder.

\subsection{Related Work}
Sparse regression codes (SPARCs) are a joint modulation and coding technique introduced by Joseph and Barron \cite{SPARC1_intro}. Unlike traditional coded modulation techniques \cite{codedMod1, codedMod2}, a codeword of SPARCs is constructed by the direct multiplication of a dictionary matrix and a sparse message vector under a block sparsity constraint. Using the Gaussian random dictionary matrix with independent and identically distributed (IID) entries for encoding and the optimal maximum likelihood (ML) decoding. SPARCs have been shown to achieve any fixed rate smaller than the capacity of Gaussian channels as the code length goes to infinity \cite{SPARC1_intro}.

Designing low-complexity decoding algorithms for SPARCs is of great interest to make the codes feasible in practice \cite{SPARC2_adaptive, SPARC3_adaptive, SPARC4, SPARC5_AMP, SPARC6_AMP}. The adaptive successive decoding method and its variations have made significant progress in this direction \cite{SPARC2_adaptive, SPARC3_adaptive}. By interpreting the decoding problem of a SPARC with $L$ sections as a multi-user detection problem in the Gaussian multiple-access channel with $L$ users under a total sum-power constraint, the idea of adaptive successive decoding is to exploit both the successive interference cancellation at the decoder with a proper power allocation strategy at the encoder. Decoding algorithms using compressed sensing have also received significant attention as computationally-efficient alternatives owing to a deep connection between SPARCs and compressed sensing \cite{CS}. In principle, the decoding problem of SPARCs can be interpreted through the lens of sparse signal recovery from noisy measurements under a certain sparsity structure. Exploiting this connection, approximate message passing (AMP) \cite{AMP}, successfully used in the sparse recovery problem, has been proposed as a computationally-efficient decoding method of SPARCs \cite{SPARC5_AMP, SPARC6_AMP}. One key feature of the AMP decoder is that the decoding performance per iteration can be analyzed by the state evolution property \cite{AMP}. Although both low-complexity decoders can decrease the error probability with a near-exponential order in the code length as long as a fixed code rate is below the capacity, the finite length performance of the AMP decoder is much better than that of the adaptive successive hard-decision decoder. However, the performance of SPARCs with such low-complexity decoders is limited in the regime of both very low rate and short-blocklength, in which a transmitter sends a few tens of information bits to a receiver using a few hundreds of the channel uses.

The performances of SPARCs in the low-rate and short-length regime can be improved by carefully designing their dictionary matrices of a finite size \cite{RIP1, RIP2}. However, finding the optimal dictionary matrix for given code rates and blocklengths is a very challenging task. To avoid this difficulty, the common approach in designing the dictionary matrix is to exploit well-known orthogonal matrices. For instance, the use of the Hadamard-based dictionary matrix is shown to provide better performances than that of the IID Gaussian random dictionary matrices in the finite-blocklength regime \cite{SPARC4}. The quasi-orthogonal sparse superposition code \cite{quasiOrtho} is another example, in which Zadoff-Chu sequences are harnessed to construct a dictionary matrix, ensuring the near-orthogonal property. Interestingly, it performs better than polar codes in some short-blocklength regimes. However, constructing such a near-orthogonal dictionary matrix per code rate and blocklength requires a high computational complexity. In addition, the decoding complexity and latency of the iterative decoder, called belief propagation successive interference cancellation, cannot meet the stringent requirements of no-error-floor performance in URLLC.

%Orthogonal sparse superposition (OSS) codes, a new class of SPARCs, have been recently suggested as a practical coding scheme for efficient short-packet communication over the additive white Gaussian noise (AWGN) channel \cite{OSS}. The OSS code differs from SPARCs in that a codeword is a sparse linear combination of orthogonal columns of a unitary dictionary matrix. Indices of these columns participating in a codeword convey information. It turns out that a short OSS code with a simple element-wise maximum a posteriori (MAP) decoder outperforms a SPARC under AMP decoding and a polar code under successive cancellation (SC) decoding. However, the OSS code uses a single unitary matrix in encoding, which results in extremely low code rates. This restriction hinders practical realizations of the OSS code. 

\subsection{Contributions}
%\begin{table}[h!]
%    \centering
%    \begin{tabular}{ | l | c | c | c |  }
%     \hline 
%     \multicolumn{4}{|c|}{\textbf{URLLC Channel Coding Candidates}} \\
%     \hline
%     & PAC & polar & BOSS \\
%     \hline
%     Low, flexible code rate & $\Circle$ & $\Circle$ & $\Circle$ \\  
%     Performance & $\circledcirc$ & $\Circle$ & $\Circle$ \\ 
%     Decoding latency & $\times$ & $\triangle$ & $\Circle$ \\ 
%     \hline
%    \end{tabular}
%    \caption{Comparison of potential URLLC coding schemes.}
%    \label{tab1:URLLC_comparison}
%\end{table}
%In this paper, we mainly consider communication over a Gaussian channel in the power-limited regime, in which the capacity $C$ approaches zero as the blocklength goes to infinity. For efficient communication in this regime, we introduce a new class of SPARCs, namely block orthogonal sparse superposition (BOSS) codes, which generalizes the encoding mechanism of the orthogonal sparse superposition (OSS) codes \cite{OSS}. The key innovation of the proposed coding scheme is to construct a codeword as a superposition of orthogonal sub-codewords, while a fat dictionary matrix consisted of multiple unitary matrices grants flexibility to coding rate. We then study the BOSS code in a fading scenario. Some preliminary results of this paper are given in \cite{BOSS1,BOSS2}. 

The major contributions of this paper are summarized as follows:

%Table \ref{tab1:URLLC_comparison} compares BOSS codes with PAC and polar codes in the context of three major aspects of URLLC channel coding. 

\begin{itemize}
    % \item We first present an encoding method  called \textit{successive encoding}. The key idea of successive encoding is to sequentially map fractions of an incoming bit stream into non-zero coefficients of sparse sub-message vectors, while their non-zero supports are chosen to be mutually exclusive. The sub-messages, then, are multiplied by a dictionary matrix to generate sub-codewords, i.e., they are a subset of weighted column vectors of the dictionary matrix. As a result, all sub-codewords are informed with the orthogonal property. The proposed encoding scheme can construct codebooks with flexible coding rates for a given blocklength by adjusting the code parameters including the dictionary matrix dimension, number of sub-codewords, sparsity level of each sub-message, and non-zero alphabets. The BOSS code has a number of intriguing aspects. Both classical permutation modulation codes introduced in the 1960s \cite{permutation} and recently introduced index modulation techniques \cite{indexMod1, indexMod2, indexMod3} can be interpreted as special cases of the BOSS code. 
    
    \item Our main contribution is to introduce a new class of sparse superposition codes, referred to as BOSS codes. Unlike a SPARC using a random dictionary matrix for encoding, the BOSS encoder exploits a structured dictionary matrix formed by the concatenation of $G$ unitary matrices of size  $M$ by $M$. The chosen $G$ unitary matrices (e.g., Fourier, Walsh-Hadamard, Haar, and discrete cosine matrices) can take a fast unitary transform for efficient encoding and decoding. The encoder constructs a codeword with this structured matrix by multiplying the dictionary matrix with a sparse message vector. A sequential bit mapping strategy generates the sparse message vector. The key idea of the sequential bit mapping is to successively map the fractions of information bits into the positions and the coefficients of sparse sub-message vectors using the non-selected column indices in the previously generated sub-message vectors. The remarkable property of our construction is that all codewords are mutually orthogonal, i.e., this is a class of orthogonal codes. In addition, it allows preserving the norms of sparse message vectors after multiplying them by the dictionary matrix. This implies that our code construction achieves the zero restricted isometry property (RIP) value from a compressive sensing viewpoint \cite{CS, RIP1}. In addition, our encoding requires linear complexity in blocklength, and it is flexible to generate various code rates for a given blocklength by adjusting the code parameters.

    % The key idea of sequential encoding is to sequentially map the fractions of information bits into non-zero coefficients of sparse sub-message vectors, while their non-zero supports are chosen to be mutually exclusive. The sub-messages, then, are multiplied by a dictionary matrix to generate sub-codewords, i.e., they are a subset of weighted column vectors of the dictionary matrix. As a result, all sub-codewords are informed with the orthogonal property.  

    %The BOSS code has a number of intriguing aspects. Both classical permutation modulation codes introduced in the 1960s \cite{permutation} and recently introduced index modulation techniques \cite{indexMod1, indexMod2, indexMod3} can be interpreted as special cases of the BOSS code. 
    
    \item We present a fast and low-complexity decoder for BOSS codes while achieving a near maximum likelihood (ML) decoding performance. We refer to this as \textit{two-stage MAP decoder}. In the first stage, the decoder takes $G$ unitary transforms of the received signal in parallel, which requires a complexity of $\mathcal{O}(GM\log (M))$. Then, with each transformed signal, the decoder independently performs element-wise MAP decoding with successive support set cancellation as in \cite{Bayesian_MP} to recover the $K$-sparse message vectors, which requires a complexity order of $\mathcal{O}(KM)$. In the second stage, using the decoded sparse message vectors in the prior stage, the decoder finds the unitary matrix index using the minimum distance detection, which requires linear complexity with the number of sub-unitary matrices, i.e., $\mathcal{O}(G)$. For a two-layered BOSS code, which is the most practically relevant case for a low-rate code design, it turns out that a simple ordered statistics (OS) decoder with linear complexity in the blocklength, $\mathcal{O}(K\log (M))$, can be optimal for the first stage decoding. The key feature of our two-stage decoder can be implemented in parallel; thereby, fast and low-complexity decoding is possible, which is paramount for extremely low-latency communications.
    
        \item  We derive an exact expression for block error rates (BLERs) of a  single-layered BOSS code with the two-stage decoding in terms of relevant code parameters, including code blocklength, code rates, and the the SNR. Our analytical expression elucidates how the code performance changes according to these code design parameters. Specifically, we confirm that the BLER performance improves as the blocklength and SNR increase, while it deteriorates with the number of unitary matrices, $G$. We also verify that the BLER performance highly depends on the first stage decoding error. This implies that increasing the code rate by using more unitary matrices is preferable in the code design at the cost of decoding complexity. Our analytical expression for BLER is also particularly useful for predicting the minimum required SNR to achieve an extremely low target BLER below $10^{-6}$, which is very hard to obtain even with computer simulations. Using this, we verify that no error-floor occurs for our decoding method as the SNR increases.

    \item To enhance coding performance, we also put forth a CRC-aided BOSS code, called a CA-BOSS code, which is constructed by serially concatenating a CRC code with a BOSS code as an outer code. For efficient decoding of CA-BOSS codes, we propose a list decoding method. The key distinction with the two-stage MAP decoder is that the list decoder finds a set of codewords with high reliabilities and validates whether they satisfy the CRC constraints. Then, in the second stage, it performs minimum distance detection for partial codewords that were successful in the validation. The decoding complexity of the list decoder does not scale up the total decoding complexity order as it only slightly increases the OS decoder complexity by a linear factor $Q$ in linear, $\mathcal{O}(QK\log (M))$. The remarkable observation is that our CA-BOSS code outperforms the CA-polar code with SC list (SCL) decoding when the block length is short, and the code rate is low. More importantly, from simulations, we verify that our code can achieve the finite-blocklength capacity within one dB using efficient encoding and decoding with the complexity order of $\mathcal{O}(GM\log (M))$.

\end{itemize}

The rest of this paper is structured as follows. Section II presents an encoding method for BOSS codes. Section III explains the approximate MAP decoder for fast decoding. Section IV provides an analytical expression for the BLER of single-layered BOSS codes when the proposed decoder is applied. Section V provides numerical results. Finally, Section VI concludes the paper with possible extensions.

% The rest of this paper is structured as follows. Section II introduces BOSS coding and presents the successive encoding method along with its variants. Section III explains the powerful two-stage MAP decoding algorithm for BOSS codes over the standard Gaussian channel. Section IV analyzes the error probability of the proposed BOSS code. Section V proposes a robust decoding method for BOSS codes in the multi-path fading channel. Section VI provides numerical results. Finally, Section VII concludes the paper with possible extensions.

\section{Block Orthogonal Sparse Superposition Coding}
In this section, we present a novel encoding strategy called successive orthogonal encoding. To augment understanding of BOSS coding, we introduce some useful properties and remarks.
\subsection{ Preliminaries}
Before describing the code construction process, we introduce some notations and definitions used in this paper.

\textbf{Additive white Gaussian noise (AWGN) channel: } We mainly investigate a transmission of codewords over the AWGN channel. We restrict our attention to the real AWGN channel, but the extension to the complex system is straightforward. Let $M \in \mathbb{Z}^+$ be the blocklength and $R \in \mathbb{R}^+$ be the rate of a code.  In the AWGN channel, when sending a codeword ${\bf c}\in \mathbb{R}^M$, the received vector $\mathbf{y} \in \mathbb{R}^M$ is obtained as 
\begin{equation}
    \mathbf{y} = \mathbf{c} + \mathbf{v},
\end{equation} where $\mathbf{v}$ is an $M$-dimensional zero-mean Gaussian noise vector of variance $\sigma_v^2$, i.e., $\mathbf{v} \sim \mathcal{N}(\mathbf{0}_M, \sigma_v^2 \mathbf{I}_M )$.

\textbf{Dictionary and sparse message vector: } A BOSS code is defined by a dictionary matrix $\mathbf{A}$ of dimension $M \times N$, where $N \in \mathbb{Z}^+$ is the length of a sparse message vector. The dictionary matrix $\mathbf{A}$ is constructed as a concatenation of $G \in \mathbb{Z}^+$ unitary matrices $\mathbf{U}_g \in \mathbb{R}^{M \times M}$:
\begin{equation}
    \mathbf{A} = \begin{bmatrix}
    \mathbf{U}_1 & \mathbf{U}_2 & \cdots & \mathbf{U}_G
    \end{bmatrix}.
\end{equation} 
It is worthwhile to note that any random unitary matrices can be used for encoding; yet a set of special unitary matrices are preferred for efficient encoding and decoding. For instance, the encoder uses a Walsh-Hadard matrix for ${\bf U}_1$.  By taking a row permutation to ${\bf U}_1$, it is possible to construct other unitary matrices as
\begin{align}
	{\bf U}_g = {\bf P}_g{\bf U}_1,
\end{align}
where ${\bf P}_g\in \{0,1\}^{M\times M}$ is the $g$th permutation matrix. To make the sub-matrices distinguishable, the encoder should choose $G-1$ distinct permutation matrices, i.e., ${\bf P}_g \neq {\bf P}_{g'}$ for $g,g'\in \{2,3,\ldots, G\}$. 

A BOSS codeword $\mathbf{c}$ is represented as a matrix-vector multiplication of the dictionary matrix and a sparse message vector ${\bf x} \in \mathbb{R}^N$, i.e., $\mathbf{c} = \mathbf{A} {\bf x}$.  The sparse message vector is generated as a superposition of $L$ layered sparse message vectors:
\begin{align}
	{\bf x} = \sum_{\ell=1}^L {\bf x}^{(\ell)},
\end{align}
where ${\bf x}^{(\ell)} \in \mathbb{R}^N$ denotes the $\ell$th layered sparse message vector with $K_\ell(\ll N)$ non-zero coefficients, i.e., $ \| {\bf x}^{(\ell)} \|_0=| \text{supp} \left({\bf x}^{(\ell)}\right) |  = K_\ell$.  The encoder assigns different signal levels for the non-zero elements in ${\bf x}^{(\ell)}$. Let $J_\ell$ be a non-zero alphabet size of ${\bf x}^{(\ell)}$. Then, the constellation for the $\ell$th sub-message vector is $\mathcal{A}_\ell = \{ \alpha_{\ell, 1}, \alpha_{\ell, 2}, \dots, \alpha_{\ell, J_\ell} \}$ such as an amplitude modulation (PAM) signal set. For simplicity in the decoding process, the signal level sets for distinct message vectors are made to be disjoint, i.e., $\mathcal{A}_j \cap \mathcal{A}_k = \emptyset$ for $j \ne k \in [L]$, where $[L]\triangleq \{1, 2, \dots, L\}$. 

%Phase-shift keying can make a good alternative set, while keeping the transmit power low, but we do not consider it in this paper. 
% 

%Throughout this paper, a log without base should be taken to be the natural logarithm. In BOSS codes, non-zero indices of ${\bf x}$ convey information. Detailed bit-mapping techniques will be explained in the following sub-section. 

% \begin{figure*}[h]
%     \centering
%     \includegraphics[width=1.0\textwidth]{BOSS_Encoder_V1.png}
%     \caption{png test}
%     % \label{Fig1:BOSS_Encoder}
% \end{figure*}

\begin{figure*}[h]
    \centering
    \includegraphics[width=1.0\textwidth]{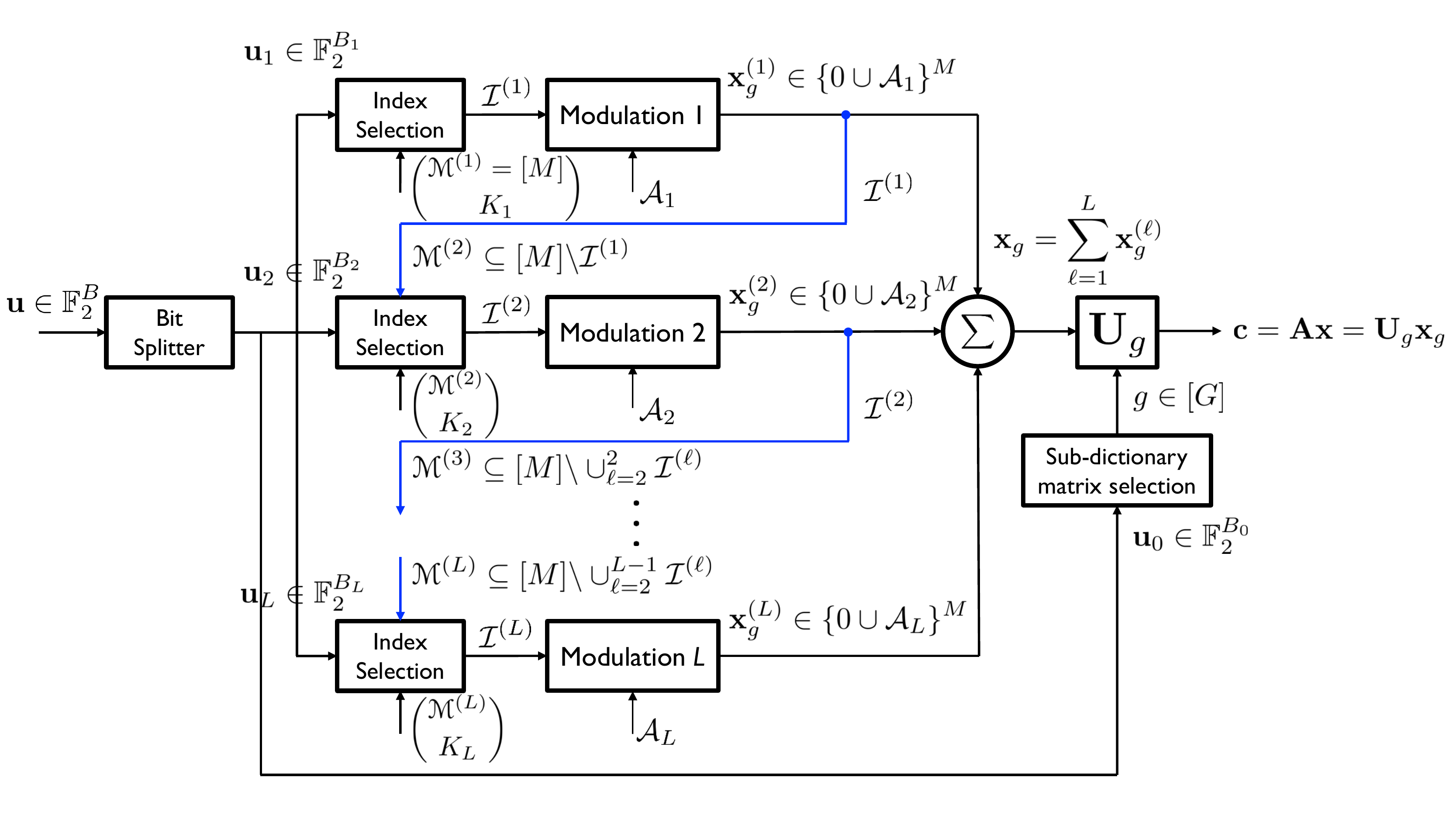}
    \caption{Proposed sequential encoder structure for the BOSS code construction.}
    \label{Fig1:BOSS_Encoder}
\end{figure*}

% \begin{figure*}[h]
%     \centering
%     \includegraphics[width=1.0\textwidth]{BOSS_Encoder.pdf}
%     \caption{Proposed sequential encoder structure for the BOSS code construction.}
%     \label{Fig1:encoder}
% \end{figure*}

\subsection{Sequential Bit Mapping}
The proposed encoding takes two stages as illustrated in Fig. \ref{Fig1:BOSS_Encoder}. In the first stage, the encoder selects a sub-dictionary matrix index $g$; $B_0 = \lfloor \log_2 ( G) \rfloor$ data bits are mapped into this block-index selection. Let us denote by ${\bf x}_g \in \mathbb{R}^M$ a segment of ${\bf x} \in \mathbb{R}^{N}$ that corresponds to the selected unitary matrix $\mathbf{U}_g$, i.e.,
\begin{equation}
    {\bf x} = \begin{bmatrix}
    {\bf x}_1^\mathsf{T} & \cdots & {\bf x}_g^\mathsf{T} & \cdots & {\bf x}_G^\mathsf{T} 
    \end{bmatrix}^\mathsf{T}.
\end{equation} Among $G$ segments of ${\bf x}$, only ${\bf x}_g$ has $K = \sum_{\ell=1}^L K_\ell$ non-zero elements, i.e., $\| {\bf x} \|_0 = \| {\bf x}_g \|_0 = K$. 

The second stage consists of $L$ successive steps. Let $\mathbf{u}_1 \in \mathbb{F}_2^{B_1}$ be a binary information string of length $B_1$. In the first layer, the encoder maps $\mathbf{u}_1$ into ${\bf x}^{(1)} \in \{ 0 \cup \mathcal{A}_1 \}^N$ by uniformly selecting $K_1$ columns of $\mathbf{U}_g$; hence, the permissible set is $\mathscr{M}^{(1)} = [M]$. The encoder, then, assigns values of $\mathcal{A}_1$ to the corresponding entries of the $g$th segment of ${\bf x}^{(1)}$, i.e., ${\bf x}_g^{(1)}$. We define the support of ${\bf x}_g^{(1)}$, i.e., a set of non-zero index locations, as $\mathcal{I}^{(1)} = \{ m \in \mathscr{M}^{(1)} | x^{(1)}_{g,m} \in \mathcal{A}_1 \}$. The first sub-codeword $\mathbf{c}^{(1)} = \mathbf{A} {\bf x}^{(1)}$ conveys $B_1 = \left\lfloor \log_2 \left( \binom{| \mathscr{M}^{(1)} | }{K_1} \right) \right\rfloor + \lfloor K_1 \log_2 ( |\mathcal{A}_1 | ) \rfloor$ information bits.

Previously chosen indices are excluded to avoid duplicate uses of locations. Utilizing the support information of ${\bf x}_g^{(1)}$, $\mathcal{I}^{(1)}$, the encoder defines a candidate set for the second layer as $\mathscr{M}^{(2)} \subseteq [M] \backslash \mathcal{I}^{(1)}$. This guarantees that $\mathcal{I}^{(1)} \cap \mathcal{I}^{(2)} = \emptyset$. The encoder then selects $K_2$ positions in $\mathscr{M}^{(2)}$ and generates ${\bf x}^{(2)}$ by allocating elements of $\mathcal{A}_2$ into the chosen positions of ${\bf x}_g^{(2)}$. The resultant sub-codeword $\mathbf{c}^{(2)} = \mathbf{A} {\bf x}^{(2)}$ contains $B_2 = \left\lfloor \log_2 \left( \binom{| \mathscr{M}^{(2)}|}{K_2} \right) \right\rfloor + \lfloor K_2 \log_2 ( | \mathcal{A}_2 | ) \rfloor$ information bits.  

The encoder successively maps original bits in the same fashion until the $L$th layer, where the last candidate set is given by $\mathscr{M}^{(L)} \subseteq [M] \backslash \cup_{j=1}^{L - 1} \mathcal{I}^{(j)}$. $\mathbf{u}_L \in \mathbb{F}_2^{B_L}$ is encoded into ${\bf x}^{(L)}$ such that $K_L$ non-zero indices of ${\bf x}_g^{(L)} \in \{0 \cup \mathcal{A}_L \}^M$ form a subset of $\mathscr{M}^{(L)}$. Thanks to the orthogonal construction, the support of ${\bf x}_g^{(L)}$, $\mathcal{I}^{(L)}$, is mutually exclusive with the union of the supports of ${\bf x}_g^{(j)}$ for $j \in \{1, 2, \dots, L - 1\}$, i.e., $\mathcal{I}^{(L)} \cap \left( \cup_{j=1}^{L - 1} \mathcal{I}^{(j)} \right) = \emptyset$.

Given that index support sets of distinct layers are non-overlapping, a BOSS codeword can be represented as a superposition of $L$ sub-codeword vectors, each of which is a linear combination of a subset of column vectors of $\mathbf{U}_g$:
\begin{equation}
    \mathbf{c} = \sum_{\ell = 1}^L \mathbf{c}^{(\ell)} = \sum_{\ell=1}^L \mathbf{A} {\bf x}^{(\ell)} = \sum_{\ell=1}^L \mathbf{U}_g {\bf x}_g^{(\ell)}.
\end{equation}

\subsection{Properties}
To shed further light on the significance of our code construction method, we provide some properties.

\textbf{Orthogonality: } The most prominent property of the BOSS code is orthogonality between sub-codewords, i.e., $ \langle \mathbf{c}^{(j)}, \mathbf{c}^{(k)} \rangle = 0$ for $j \ne k \in [L]$. This inherent orthogonal property helps develop a computationally efficient yet powerful decoder, which will be explained in Section III.

\textbf{Zero-RIP codebook: } In the sparse recovery literature, RIP constants measure change in the $\ell_2$ norm of sparse vectors induced by the dictionary matrix and therefore are a popular metric to analyze the quality of sparse recovery algorithms \cite{RIP1, RIP3}. Let us denote by $\mathscr{X}$ a set of possible sparse message vectors
\begin{equation}
    \mathscr{X} = \left\{ {\bf x} | {\bf x} \in \mathcal{A}^N , \| {\bf x} \|_0 = \sum_{\ell=1}^L K_\ell , \sum_{g=1}^G \mathbf{1}_{\{ \| {\bf x}_g \|_0 \ne 0 \}} = 1 \right\}, \label{msgVectorSet}
\end{equation} where $\mathcal{A} = \{0\} \bigcup \left( \cup_{\ell=1}^L \mathcal{A}_\ell \right)$. Entailed by the proposed code construction method, BOSS codewords constitute a zero-RIP codebook over $\mathscr{X}$:
\begin{equation}
    \mathscr{C} = \{ \mathbf{c} | \mathbf{c} = \mathbf{A} {\bf x} \},
\end{equation} where ${\bf x} \in \mathscr{X}$, and $\| \mathbf{c} \|_2 = \| \mathbf{A} {\bf x}  \|_2 = 0$.  This property elucidates the difference with SPARCs. For encoding of SPARCs, the Gaussian random dictionary matrix is used with a random sparse message vector. Therefore, in SPARCs, the norm of codewords guarantees $(1-\delta)\| \mathbf{c} \|_2\leq   \| \mathbf{A} {\bf x}  \|_2  \leq (1+\delta) \|\mathbf{c} \|_2$ with RIP constant $\delta$ \cite{RIP1}.

 %Therefore, the proposed BOSS code is more beneficial than SPARCs in a low-rate and short-blocklength regime

%This property puts the BOSS code on a sound theoretical footing, and successful recovery of the sparse vector ${\bf x}$ is expected.

\textbf{Decodability: } In the absence of noise, a single-block, i.e., $G = 1$, BOSS code is uniquely decodable since $\mathcal{A}_j \cap \mathcal{A}_k = \emptyset$ for $j \ne k \in [L]$. Suppose $\mathbf{A} = \mathbf{I}_M$, i.e., $\mathbf{c} = {\bf x}$. This is true because a decoder is able to disnguish the $\ell$th sub-codeword ${\bf x}^{(\ell)}$ from ${\bf x} = \sum_{j = 1}^L {\bf x}^{(j)}$, provided $x^{(\ell)}_m \notin \cup_{j \ne \ell} \mathcal{A}_j, \, \forall m \in [M]$. The decoder, then, performs de-mapping from ${\bf x}^{(\ell)}$ to $\mathbf{u}_\ell$, a stream of $B_\ell$ information bits.

\textbf{Code rate: } At the $\ell$th layer, the encoder is allowed the leeway to choose $K_\ell$ indices from $\mathscr{M}^{(\ell)}$; thereby, it maps $B_\ell = \left\lfloor \log_2 \left( \binom{|\mathscr{M}^{(\ell)}|}{K_\ell} \right) \right\rfloor + \lfloor K_\ell \log_2 ( | \mathcal{A}_\ell | ) \rfloor$ information bits. Taking account of $B_0$ bits mapped in the first encoding stage, the rate of the BOSS code is
\begin{equation}
    R = \frac{\lfloor \log_2 (G) \rfloor + \sum_{\ell=1}^L \left( \left\lfloor \log_2 \left( \binom{|\mathscr{M}^{(\ell)}|}{K_\ell} \right) \right\rfloor + \lfloor K_\ell \log_2 ( | \mathcal{A}_\ell | ) \rfloor \right) }{M}.
\end{equation} For a symmetric case in which $| \mathscr{M}^{(\ell)} | = \mathcal{M}(< M)$, $K_\ell = K$, and $J_\ell = 1$ for all $\ell \in [L]$, the code rate is simplified to $R = \frac{\lfloor \log_2 (G) \rfloor + L \left\lfloor \log_2 \left( \binom{\mathcal{M}}{K} \right) \right\rfloor}{M}$. For a fixed blocklength $M$, the proposed encoding scheme can construct codes of various rates by appropriately tuning multiple design parameters: the number of blocks $G$, the layer depth $L$, the number of non-zeros per layer $K_\ell$, and the size of non-zero alphabets $J_\ell$. This flexibility in coding rate is a salient feature of BOSS coding and attests to its applicability to URLLC use-cases.

\textbf{Average transmit power: } One intriguing property of the BOSS code is that its average transmit power is tiny, thanks to the sparsity in the code construction. Without loss of generality, we assume that the encoder has chosen the very first sub-dictionary matrix, i.e., $\| {\bf x} \|_0 = \| {\bf x}_1 \|_0 = K$. The $\ell$th layer's vector ${\bf x}_1^{(\ell)}$ has $K_\ell$ elements drawn from $\mathcal{A}_\ell$, and the signal energy associated with few non-zero entries is distributed by $\mathbf{U}_1$. Since the norm is preserved under unitary transformation, the average power of a sub-codeword $\mathbf{c}^{(\ell)}$ is given by
\begin{equation}
    \mathbb{E} \{ \| \mathbf{c}^{(\ell)} \|_2^2 \} = \mathbb{E} \{ \| \mathbf{U}_1 {\bf x}_1^{(\ell)} \|_2^2 \} = \mathbb{E}\{ \| {\bf x}_1^{(\ell)} \|_2^2 \} = \frac{K_\ell \frac{\sum_{j=1}^{J_\ell} \alpha_{\ell, j}^2}{J_\ell}}{M} .
\end{equation} Since all sub-codeword vectors are orthogonal, the average transmit power becomes
\begin{equation}
    E_{\rm{s}}= \mathbb{E} \{ \| \mathbf{A} {\bf x} \|_2^2 \} = \sum_{\ell = 1}^L \mathbb{E} \{ \| {\bf x}_1^{(\ell)} \|_2^2 \} = \sum_{\ell = 1}^L \frac{K_\ell \sum_{j=1}^{J_\ell} \alpha_{\ell, j}^2}{J_\ell M}.
\end{equation}

%\begin{figure*}[h]
%    \centering
%\includegraphics[width=0.6\textwidth]{BOSS_MappingTable.png}
%    \caption{Bit-mapping table of the second layer.}
%    \label{Fig2:Table}
%\end{figure*}

\textbf{Example: } For ease of exposition, let us consider a BOSS code with $[G, M, L] = [8, 64, 2]$, $K_1 = K_2 = 1$, and singleton PAM alphabets $\mathcal{A}_1 = \{ 1 \}$ and $\mathcal{A}_2 = \{ -1 \}$. The first $B_0 = \lfloor \log_2 (8) \rfloor = 3$ bits are encoded into a block index $g \in [8]$. The encoder maps the following $B_1 = \lfloor \log_2 \left( \binom{64}{1} \right) \rfloor = 6$ bits into choosing a column index $i_1 \in [64]$. Without loss of generality, suppose $g = 1$ and $i_1 = 1$. Then, ${\bf x}_1^{(1)}$ has a single non-zero coefficient, i.e., $x_1^{(1)} = 1$. The encoder chooses a second position $i_2$ from $\mathscr{M}^{(2)} = [M] \backslash \{ i_1 \} = \{2, 3, \dots, 64\}$ represented by the last $B_2 = \left\lfloor \log_2 \left( \binom{63}{1} \right) \right\rfloor = 5$ bits based on the predefined mapping table, and assigns $-1$ to the selected entry. The codeword is then a linear combination of the selected columns of $\mathbf{U}_1$, i.e., $\mathbf{c} = \mathbf{U}_1 \left( {\bf x}_1^{(1)} + {\bf x}_1^{(2)} \right) = \mathbf{u}_{1, i_1} - \mathbf{u}_{1, i_2}$, where $\mathbf{u}_{1,j}$ is a $j$th column vector of $\mathbf{U}_1$ for $j \in [M]$.  A codeword delivers a total of $14=(3+6+5)$ bits with $64$ channels uses, i.e., $R=\frac{14}{64}=0.21875$. The normalized average transmit power per channel use becomes $E_s = \frac{1}{64} + \frac{1}{64} = \frac{1}{32}$.

\subsection{Remarks}
We provide some remarks to offer further insights into BOSS coding and highlight the difference with the existing coding and modulation methods.

\textbf{Remark 1 (Joint encoding): } The block-index selection and first-layer bit-mapping can be combined into a single process. The encoder chooses a single column index from a fat dictionary matrix, i.e., $\binom{N}{1}=\binom{G}{1}\binom{M}{1}$, and the remaining $K_1 - 1$ indices are selected from a corresponding block. That is, putting the alphabet allocation aside,
\begin{equation}
    \log_2 (G) + \log_2 \left( \binom{M}{K_1} \right) = \log_2 (N) + \log_2 \left( \binom{M - 1}{K_1 - 1} \cdot \frac{1}{K_1} \right).
\end{equation}

\textbf{Remark 2 (Orthogonal multiplexing for multi-layer index modulation signals): } The proposed coding scheme also generalizes the existing index modulation methods \cite{indexMod1}. Consider BOSS encoding with $G = 1$ and $L = 1$. The resulting code is identical to the index modulation. Therefore, our coding scheme can be interpreted as an efficient orthogonal multiplexing method of multi-layer index (spatial) modulated signals \cite{indexMod1, indexMod2, indexMod3}.  

%\textbf{Remark 3 (Difference with permutation modulation codes): } One interesting connection is that classical permutation modulation codes are a special case of our BOSS codes with the joint information bit-mapping technique. Specifically, by using a single unitary matrix and setting signal legvel sets to be a singleton $\mathcal{A}_\ell = \{ \alpha_{\ell} \}$ for $\ell \in [L]$, it is possible to generate a codebook indetical to Variant I in \cite{permutation} with a joint mapping method, which achieves the rate of
%\begin{equation}
%    R = \frac{\left\lfloor \log_2 \left( \prod_{\ell = 1}^L \binom{M - (\ell - 1)K}{K} \right) \right\rfloor}{M}. 
%\end{equation} Thanks to additional bits mapped into the non-zero positions and great latitude in designing the constellations per layer, our encoding scheme is able to generate a large codebook for given $M$ and $K$. Besides, the superposition encoding method of multiple sub-codewords facilitates implementation of the encoder in practice because it greatly reduces the encoding complexity by the separate bit-mapping technique. 

\begin{figure*}[h]
    \centering
    \includegraphics[width=0.8\textwidth]{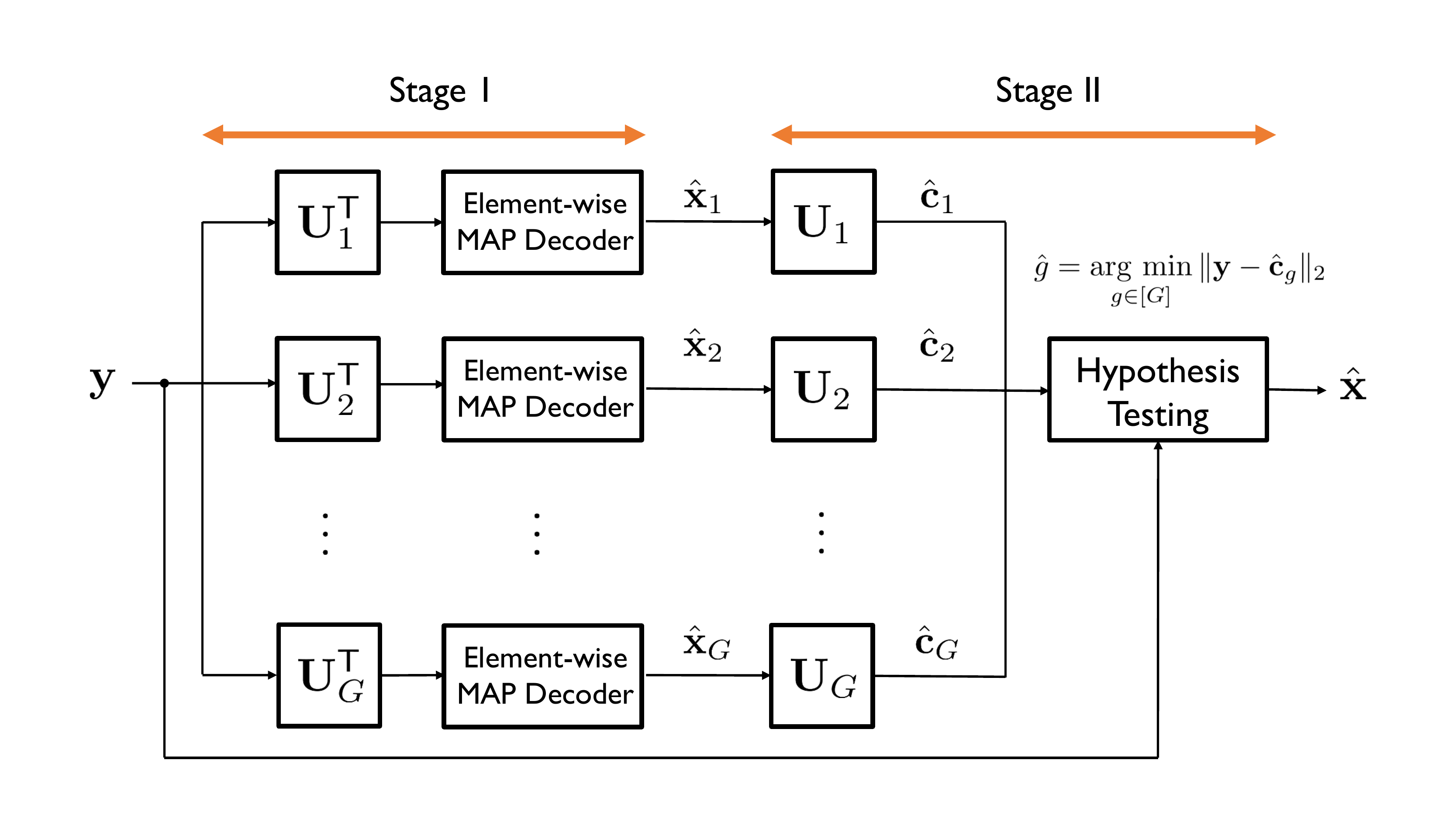}
    \caption{Two-stage MAP decoder for BOSS codes in the AWGN channel.}
    \label{Fig2:BOSS_Decoder}
\end{figure*}

\section{Approximate MAP Decoder}
This section presents an approximate MAP decoder featuring low decoding latency and complexity. We first explain the motivation of the approximate MAP decoder and present the two-stage decoding algorithm in the sequel. 

\subsection{Two-stage MAP Approximation}  Under the premise that a codeword ${\bf c}$ is generated using the $g$th sub-dictionary matrix in ${\bf A}$, the codeword can be written as
\begin{align}
	{\bf c}={\bf A}{\bf x}={\bf U}_g{\bf x}_g,
\end{align}
where the sub-message vector ${\bf x}_g = \sum_{\ell=1}^L {\bf x}_g^{(\ell)}$ is a member of the set
\begin{equation}
    \mathscr{X}_g = \left\{ {\bf x}_g | {\bf x}_g \in \mathcal{A}^M, \| {\bf x}_g \|_0 = \sum_{\ell=1}^L K_\ell \right\} \label{sub-signal set}.
\end{equation} 
Let  $\mathscr{H}_g$ be a hypothesis that the transmitted codeword is constructed with a sub-dictionary matrix $\mathbf{U}_g$:
\begin{equation}
    \mathscr{H}_g : \mathbf{c} = \mathbf{U}_g {\bf x}_g.
\end{equation} 
Then, the MAP decoding problem can be decomposed into two different sub-MAP tasks, i.e., 
\begin{align*}
    \hat{{\bf x}}^{\text{MAP}} &= \underset{{\bf x} \in \mathscr{X}}{\arg \, \max} \, \mathbb{P} ( {\bf x} | \mathbf{y} ) \\
    &= \underset{{\bf x}_g \in \mathscr{X}_g, g \in [G]}{\arg \, \max} \, \mathbb{P} ( {\bf x}_g, \mathscr{H}_g | \mathbf{y} ) \\
    &= \underset{{\bf x}_g \in \mathscr{X}_g, g \in [G]}{\arg \, \max} \, \mathbb{P} ( {\bf x}_g | \mathbf{y}, \mathscr{H}_g ) \mathbb{P} ( \mathscr{H}_g | \mathbf{y} ). \numberthis{\label{eq:MAPdecoding}}
\end{align*}
Solving this MAP decoding problem requires the complexity order of $\mathcal{O}(|\mathscr{X}|)$, which is prohibitively high as the number of information bits increases. To diminish the decoding complexity, we present a two-stage MAP decoder.  The key idea of the two-stage MAP decoder is to solve the MAP decoding problem in \eqref{eq:MAPdecoding} independently for ${\bf x}_g \in \mathscr{X}_g$ and $g \in [G]$, respectively. To be specific, in the first stage, the decoder first identifies ${\bf \hat x}_g $ under  $\mathscr{H}_g$ 
\begin{align}
	{\bf \hat x}_g =\underset{{\bf x}_g \in \mathscr{X}_g}{\arg \, \max} \, \mathbb{P} ( {\bf x}_g | \mathbf{y}, \mathscr{H}_g ).
\end{align}
Then, in the second stage, using the decoded ${\bf \hat x}_g$, it finds the block index ${\hat g}$:
\begin{align}
	{\hat g} =\underset{g \in [G]}{\arg \, \max} \, \mathbb{P} ( \mathscr{H}_g | \mathbf{y} ).
\end{align}

\subsection{Two-stage MAP Decoder}
We explain our two-stage MAP decoding algorithm.

\textbf{Sparse message vector recovery: } As illustrated in Fig. \ref{Fig2:BOSS_Decoder}, in the first stage, the decoder aims at identifying ${\bf x}_g$ under each hypothesis $\mathscr{H}_g$ in parallel. To this end, the decoder simply takes a fast unitary transform to the received signal vector ${\bf y}$ and produces $G$ transformed receive vectors as \begin{equation}
    \mathbf{y}_g = \mathbf{U}_g^\mathsf{T} \mathbf{y} = {\bf x}_g +\Tilde{\mathbf{v}}_g,
\end{equation}
 for $g\in [G]$, and $\Tilde{\mathbf{v}}_g = \mathbf{U}_g^\mathsf{T} \mathbf{v}$ follows the identical distribution as $\mathbf{v}$ because the Gaussian distribution is invariant under the unitary transform. The complexity order to obtain ${\bf y}_g$ for $g\in [G]$ is $\mathcal{O}(G M\log M)$ thanks to the fast transform for $\mathbf{U}_g^\mathsf{T}={\bf U}_1^{\top} {\bf P}_g^{\top}$.

Now, the decoder solves the first sub-MAP decoding problem using $\mathbf{y}_g$:
\begin{equation}
    \hat{{\bf x}}_g = \underset{{\bf x}_g \in \mathscr{X}_g}{\arg \, \max} \, \mathbb{P} \left( {\bf x}_g | \mathbf{y}, \mathscr{H}_g \right) = \underset{{\bf x}_g \in \mathscr{X}_g}{\arg \, \max} \, \mathbb{P} \left( {\bf x}_g | \mathbf{y}_g \right). \label{firstStageMAP}
\end{equation} 
 Exploiting the fact that the support of ${\bf x}_g$ is a union of mutual-exclusive index sets from $L$ layers, we compute the log of joint a posteriori probability (APP) in \eqref{firstStageMAP} and factorizing:
\begin{align*}
    \log \mathbb{P} ( {\bf x}_g | \mathbf{y}_g ) &= \sum_{\ell=1}^L \log \mathbb{P} ({\bf x}_g^{(\ell)} | \mathbf{y}_g, {\bf x}^{(\ell-1)}_g, \dots , {\bf x}_g^{(2)}, {\bf x}_g^{(1)} ) \\
    &= \sum_{\ell=1}^L \log \mathbb{P} ( {\bf x}_g^{(\ell)} | \mathbf{y}_g, \mathcal{I}_g^{(\ell - 1)}, \dots , \mathcal{I}_g^{(2)}, \mathcal{I}_g^{(1)} ), \numberthis{\label{eq:APP}}
\end{align*} where the second equality follows from that previous non-zero supports, $\mathcal{I}^{(\ell-1)}_g, \dots, \mathcal{I}_g^{(2)}, \mathcal{I}_g^{(1)}$, provide sufficient information to decode ${\bf x}_g^{(\ell)}$. Hence, the decoder shall leverage side information obtained in the preceding layers to recover ${\bf x}_g^{(\ell)}$ and update the conditional likelihoods accordingly. This motivates to design a decoding algorithm based on successive support set cancellation. From the Bayes's rule as in \cite{Bayesian_MP, MAPsupp}, the $\ell$th layer's log likelihood in \eqref{eq:APP} is reformulated as
\begin{equation}
    \log \mathbb{P} ( {\bf x}_g^{(\ell)} | \mathbf{y}_g,  \hat{\mathcal{I}}^{(\ell-1)}_g, \dots, \hat{\mathcal{I}}^{(2)}_g, \hat{\mathcal{I}}_g^{(1)}) = \frac{1}{C} \sum_{m \in \hat{\mathscr{M}}_g^{(\ell)}} \log \frac{\mathbb{P} ( y_{g,m} | x^{(\ell)}_{g,m} ) \mathbb{P} ( x^{(\ell)}_{g,m} )}{\mathbb{P} (y_{g,m})} \mathbf{1}_{\{ \| {\bf x}_g^{(\ell)} \|_0 = K_\ell \}}, \label{log_likelihood}
\end{equation} where $C \in \mathbb{R}^+$ is a normalizing constant, and a candidate set estimate $\hat{\mathscr{M}}^{(\ell)}_g$ is defined as 
\begin{equation}
    \hat{\mathscr{M}}^{(\ell)}_g = [M] \backslash \cup_{j=1}^{\ell - 1} \hat{\mathcal{I}}_g^{(j)}. \label{candidate_estimate}
\end{equation} Here and hereinafter, we omit support estimates obtained beforehand for notational simplicity in \eqref{log_likelihood}, as they are already manifested in $\hat{\mathscr{M}}_g^{(\ell)}$. The permissible set in \eqref{candidate_estimate} is updated at each iteration so that prior information on previous guesses is properly incorporated into subsequent decoding processes. Note that estimates in \eqref{candidate_estimate} may differ by hypothesis $\mathscr{H}_g$, but they all feature the same cardinality predefined in encoding:
\begin{equation}
    | \hat{\mathscr{M}}^{(\ell)}_g | =  2^{B_\ell} \leq \left( M - \sum_{j=1}^{\ell - 1} K_j \right), \quad \quad \forall g \in [G].
\end{equation} The element-wise conditional likelihood function in \eqref{log_likelihood} is given by
\begin{equation}
    \mathbb{P} \left( y_{g,m} | x^{(\ell)}_{g,m} \right) = \begin{cases}
    \frac{1}{J_\ell} \sum_{j=1}^{J_\ell} \frac{1}{\sqrt{2\pi \sigma_v^2}} e^{-\frac{|y_{g,m} - \alpha_{\ell, j}|^2}{2\sigma_v^2}} & \text{if } \, x^{(\ell)}_{g,m} \in \mathcal{A}_\ell \\
    \frac{1}{\sqrt{2\pi \sigma_v^2}} e^{-\frac{| y_{g,m} |^2}{2\sigma_v^2}} & \text{else.}
    \end{cases} \label{AWGN_APP}
\end{equation} As values of $\mathcal{A}_\ell$ are uniformly allocated to non-zero entries of ${\bf x}^{(\ell)}_g$, a prior distribution of $x^{(\ell)}_{g,m}$ is given by
\begin{equation}
    \mathbb{P} ( x^{(\ell)}_{g,m} ) = \begin{cases}
    p^{(\ell)}_{g,m} ,& \text{if } x^{(\ell)}_{g,m} \in \mathcal{A}_\ell \\
    1 - p^{(\ell)}_{g,m} ,& \text{if } x^{(\ell)}_{g,m} = 0 ,
    \end{cases} \label{msgPriorDist}
\end{equation} where $p^{(\ell)}_{g,m}$ is the probability of $x^{(\ell)}_{g,m}$ being a non-zero element given by

\begin{figure*}[h]
    \centering
    \includegraphics[width=0.8\textwidth]{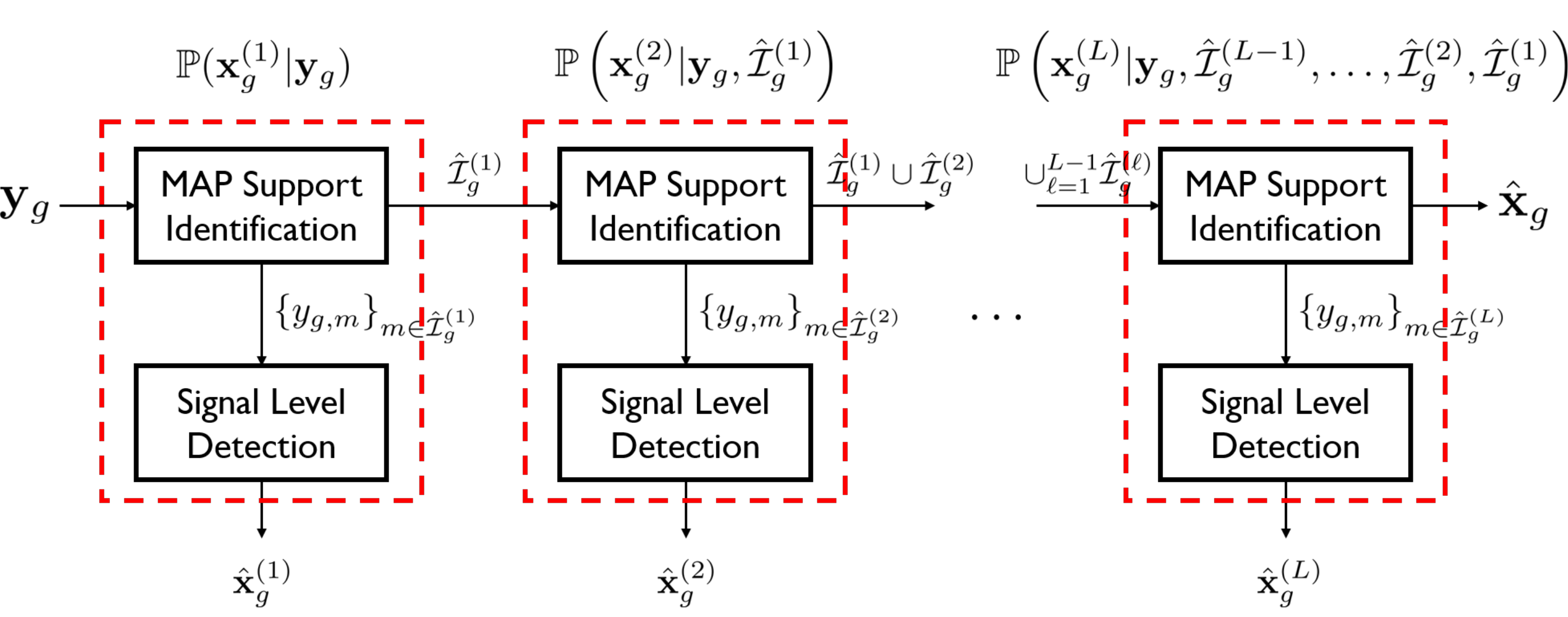}
    \caption{Recovery of a sparse message vector under hypothesis $\mathscr{H}_g$.}
    \label{Fig3:BOSS_Decoder_First_Stage}
\end{figure*}

% \begin{equation}
%     \mathbb{P} ( x^{(\ell)}_{g,m} ) = (1 - p^{(\ell)}_{g,m}) \delta (x^{(\ell)}_{g,m}) + \sum_{i=1}^{J_\ell} p^{(\ell)}_{g,m} \left(1 - \delta(x^{(\ell)}_{g,m} - \alpha_{\ell, i} ) \right),
% \end{equation} where $p^{(\ell)}_{g,m}$ is the probability of $x^{(\ell)}_{g,m}$ being a non-zero element given by

\begin{equation}
    p^{(\ell)}_{g,m} = \frac{K_\ell}{| \mathscr{M}^{(\ell)} |} = \frac{K_\ell}{M - \sum_{j=1}^{\ell - 1} K_j}. \label{uniform_probability}
\end{equation} Invoking \eqref{AWGN_APP} and \eqref{msgPriorDist}, we have the marginal distribution of $y_{g,m}$:
\begin{align*}
    \mathbb{P} ( y_{g,m} ) &= \mathbb{P} ( y_{g,m} | x^{(\ell)}_{g,m} \in \mathcal{A}_\ell) \mathbb{P} (x^{(\ell)}_{g,m} \in \mathcal{A}_\ell ) + \mathbb{P} ( y_{g,m} | x^{(\ell)}_{g,m} = 0 ) \mathbb{P} ( x^{(\ell)}_{g,m} = 0 ) \\
    &= \frac{1}{J_\ell} \sum_{j=1}^{J_\ell} \frac{1}{\sqrt{2\pi \sigma_v^2}} e^{-\frac{|y_{g,m} - \alpha_{\ell, j}|^2}{2\sigma_v^2}} p^{(\ell)}_{g,m} + \frac{1}{\sqrt{2\pi \sigma_v^2}} e^{-\frac{|y_{g,m}|^2}{2\sigma_v^2}} (1 - p^{(\ell)}_{g,m}). \numberthis{\label{}}
\end{align*}
Note that recovery of a sparse vector ${\bf x}_g^{(\ell)}$ is equivalent to identification of its non-zero element locations such that paired columns of $\mathbf{U}_g$ constitute a sub-codeword $\mathbf{c}^{(\ell)}$. For every index $m \in \hat{\mathscr{M}}^{(\ell)}_g$, the decoder computes its probability of being an element of $\mathcal{I}^{(\ell)}$, i.e. a likelihood of $x^{(\ell)}_{g,m}$ taking on value from $\mathcal{A}_\ell$ conditioned on $y_{g,m}$:
\begin{align*}
    \log \mathbb{P} (m \in \mathcal{I}^{(\ell)} | y_{g,m} )
    %&= \log \mathbb{P} (x^{(\ell)}_{g,m} \in \mathcal{A}_\ell | y_{g,m} ) \\
    &= \log \frac{\mathbb{P} ( y_{g,m} | x^{(\ell)}_{g,m} \in \mathcal{A}_\ell ) \mathbb{P} ( x^{(\ell)}_{g,m} \in \mathcal{A}_\ell )}{ \mathbb{P} (y_{g,m})}\\
    &= \log \frac{\frac{1}{J_\ell} \sum_{j=1}^{J_\ell} \frac{1}{\sqrt{2\pi \sigma_v^2}} e^{-\frac{|y_{g,m} - \alpha_{\ell, j}|^2}{2\sigma_v^2}} p^{(\ell)}_{g,m}}{\frac{1}{J_\ell} \sum_{j=1}^{J_\ell} \frac{1}{\sqrt{2\pi \sigma_v^2}} e^{-\frac{|y_{g,m} - \alpha_{\ell, j}|^2}{2\sigma_v^2}} p^{(\ell)}_{g,m} + \frac{1}{\sqrt{2\pi \sigma_v^2}} e^{-\frac{|y_{g,m}|^2}{2\sigma_v^2}} (1 - p^{(\ell)}_{g,m})}. \numberthis{\label{MAP metric}}
\end{align*} The decoder sorts $\hat{\mathscr{M}}^{(\ell)}_g$ by the MAP metric in \eqref{MAP metric}, which is monotone increasing with respect to $y_{g,m}$. To satisfy the sparsity requirement $\mathbf{1}_{\{ \|{\bf x}^{(\ell)}_g \|_0 = K_\ell \}}$ in \eqref{log_likelihood}, the decoder selects $K_\ell$ indices that have the highest likelihoods and generates an ordered support estimate $\hat{\mathcal{I}}^{(\ell)}_g$:
\begin{equation}
    \hat{\mathcal{I}}^{(\ell)}_g = \left\{ \hat{i}^{(\ell)}_{g, 1}, \hat{i}^{(\ell)}_{g, 2}, \dots, \hat{i}^{(\ell)}_{g, K_\ell} \right\},
\end{equation} where $\mathbb{P} ( \hat{i}^{(\ell)}_{g,j} \in \mathcal{I}^{(\ell)} | y_{g,j} ) \geq  \mathbb{P} ( \hat{i}^{(\ell)}_{g,k} \in \mathcal{I}^{(\ell)} | y_{g,k} )$ for $j < k$. Once the layer support estimate is determined, the decoder performs another MAP estimation to identify the signal levels of $x^{(\ell)}_{g, m}$ for $m \in \hat{\mathcal{I}}_g^{(\ell)}$. In particular, this simplifies to the minimum Euclidean distance decoding:
\begin{align*}
    \hat{x}^{(\ell)}_{g, m}
    %&= \underset{\alpha_{\ell, j} \in \mathcal{A}_\ell}{{\arg \, \max}} \, \mathbb{P} \left( x^{(\ell)}_{g,m} = \alpha_{\ell, j} | m \in \hat{\mathcal{I}}_g^{(\ell)}, y_{g,m} \right) \\
    &= \underset{\alpha_{\ell, j} \in \mathcal{A}_\ell}{{\arg \, \min}} \, | y_{g,m} - \alpha_{\ell, j} | . \numberthis{\label{AWGN_signalDetection}}
\end{align*} After $L$ iterations, we obtain $\hat{{\bf x}}_g$ whose support is $\hat{\mathcal{I}}_g = \bigcup_{\ell=1}^L \hat{\mathcal{I}}_g^{(\ell)}$. Fig. \ref{Fig3:BOSS_Decoder_First_Stage} describes the first stage under hypothesis $\mathscr{H}_g$.

\textbf{The block index recovery:} Using $\hat{{\bf x}}_g$ for $g\in [G]$, in the second stage, the decoder performs the hypothesis testing to identify a true block index $g$. For every $\hat{{\bf x}}_g$, the decoder generates a tentative codeword as $\hat{\mathbf{c}}_g = \mathbf{U}_g \hat{{\bf x}}_g$. Since $\mathbf{v}$ is the Gaussian white noise vector, and the encoder has selected the block index uniformly from $[G]$, the second sub-MAP decoding problem is reduced to seeking $\hat{\mathbf{c}}_g$ that is closest to $\mathbf{y}$:
\begin{equation}
    \hat{g} = \underset{g \in [G]}{\arg \, \max} \, \mathbb{P} ( \mathscr{H}_g | \mathbf{y} ) = \underset{g \in [G]}{\arg \, \min} \, \| \mathbf{y} - \hat{\mathbf{c}}_g \|_2 .
\end{equation} As a result, the decoder obtains the estimate of a sparse message vector of dimension $N$ with $K = \sum_{\ell = 1}^L K_\ell$ non-zero entries, all located in the $\hat{g}$th segment, as
\begin{equation}
    \hat{{\bf x}}^{\text{MAP}} = \begin{bmatrix}
    \mathbf{0}_M^\mathsf{T} & \cdots & {\hat{{\bf x}}_{\hat{g}}}^\mathsf{T} & \cdots & \mathbf{0}_M^\mathsf{T}
    \end{bmatrix}^\mathsf{T} .
\end{equation} 

% \subsection{Remarks}

\textbf{Remark 3 (Decoding complexity): } The decoding complexity order of the proposed two-stage MAP decoder is sub-quadratic in the blocklength, i.e., $\mathcal{O}(GM\log M)$. Specifically, under the hypothesis of $\mathscr{H}_g$, the decoder first computes $\mathbf{y}_g$. To accomplish this, the decoder first takes the permutation using ${\bf P}_g$ and then performs the fast transform with ${\bf U}_1$, i.e., ${\bf U}_g^{\top}={\bf U}_1^{\top}{\bf P}_g^{\top}$. This procedure requires the complexity of $\mathcal{O}(GM\log M)$. The decoder, then, calculates $| \mathscr{M}^{(\ell)} | \leq M - \sum_{j=1}^{\ell - 1} K_j$ APPs in \eqref{MAP metric} and selects $K_\ell$ indices with the largest APP values to estimate the support. Hence, $\hat{\mathcal{I}}_g^{(\ell)}$ can be obtained with the complexity order $\mathcal{O} \left( \left(M - \sum_{j = 1}^{\ell - 1} K_j \right) \log_2 ( K_\ell ) \right)$. After $\hat{\mathcal{I}}_g^{(\ell)}$ is found, the decoder performs element-wise maximum likelihood signal level detection for the $\ell$th layer, which takes $J_\ell K_\ell$ computations. The $G$ sub-message estimates $\hat{{\bf x}}_g$ are obtained by repeatedly performing the identical procedures. In the last stage, the decoder generates $G$ codeword candidates and compares their Euclidean distance from $\mathbf{y}$. This re-encoding task, however, makes a negligible contribution to the overall decoding complexity, as each candidate is a linear combination of columns of $\mathbf{U}_g$. As a result, the total decoding complexity is $\mathcal{O} \left( G M\log (M) \right)$, which is sub-quadratic in the blocklength. 

%As the complexity is governed by computation of $\mathbf{y}_g$, it can be greatly reduced to $\mathcal{O} ( G M \log_2 (M) )$ if we generate a dictionary matrix with a special unitary matrix that allows fast transform as mentioned in Remark 1.
%
%For a special case of BOSS codes with a single block, i.e., $G = 1$ and $N = M$, the decoding complexity is linear in the blocklength. A single-block BOSS code can be generated with the identity matrix, i.e., $\mathbf{A} = \mathbf{I}_M$, and decorrelation is no longer needed. Then, $\sum_{\ell = 1}^L \left[ \left(M - \sum_{j=1}^{\ell - 1} K_j \right) \log_2 ( K_\ell) + J_\ell K_\ell \right]$ computations are required to recover $\hat{{\bf x}}$. On the premise that $M \gg K_\ell$ for $\ell \in [L]$, the decoding complexity roughly becomes $\mathcal{O}(L M)$ which is linear in both the blocklength and layer depth.

%\textbf{Remark 7 (Implementation aspects of BOSS codes): } Although hardware implementation is not within the scope of this paper, we discuss some favorable aspects of BOSS coding that are appealing to practical applications. For example, the process of obtaining $\mathbf{y}_g$ can be parallelized for $G$ hypotheses, thereby significantly decreasing the required number of clock cycles. Moreover, if a Hadamard-based encoder is used, the system needs a very small amount of memory to store the dictionary matrix and row-swapping patterns.

\textbf{Remark 4 (Optimality of a simple ordered statistics decoder): }
 We consider a two-layered BOSS code with uniform sparsity $K_1 = K_2$; $\mathcal{A}_1 = \{ 1 \}$ and $\mathcal{A}_2 = \{ -1 \}$; and an arbitrary block size $G$. In this case, we show that the first procedure of the proposed two-stage MAP decoding algorithm is equivalent to a simple ordered statistics (OS) decoder. By plugging
\begin{equation}
    \mathbb{P} ( y_{g,m} | m \in \mathcal{I}^{(1)} ) =\frac{1}{\sqrt{2 \pi \sigma_v^2}} \exp \left( - \frac{(y_{g,m} - 1)^2}{2\sigma_v^2} \right)
\end{equation} and
\begin{align*}
    & \mathbb{P} ( y_{g,m} | m \notin \mathcal{I}^{(1)} )\nonumber\\
    &= \mathbb{P} (y_{g,m} | m \in \mathcal{I}^{(2)} ) \mathbb{P} ( m \in \mathcal{I}^{(2)} ) +  \mathbb{P} \left( y_{g,m} | m \notin ( \mathcal{I}^{(1)} \cup \mathcal{I}^{(2)}  ) \right) \mathbb{P} ( m \notin ( \mathcal{I}^{(1)} \cup \mathcal{I}^{(2)} ) ) \\
    & = \frac{1}{\sqrt{2\pi \sigma_v^2}} \exp \left( - \frac{(y_{g,m} + 1)^2}{2\sigma_v^2} \right) \frac{K_1 }{N - K_1}   + \frac{1}{\sqrt{2\pi \sigma_v^2}} \exp \left( - \frac{-y_{g,m}^2}{2\sigma_v^2} \right) \frac{N - 2K_1}{N} \numberthis{\label{}}
\end{align*} into \eqref{AWGN_APP}, we obtain the log APP of an event $m \in \mathcal{I}^{(1)}$:
\begin{equation}
    \log \mathbb{P} ( m \in \mathcal{I}^{(1)} | y_{g,m}) = \frac{1}{1 + \exp \left( -\frac{2 y_{g,m}}{\sigma_v^2} \right) \frac{N}{N - K_1} + \exp \left( -\frac{2y_{g,m} - 1}{2\sigma_v^2} \right)\frac{(N - 2K_1)}{K_1}}.
\end{equation} $\log \mathbb{P} ( m \in \mathcal{I}^{(1)} | y_{g,m})$ is a monotonically increasing function of $y_{g,m}$, so we conclude that the support estimate $\hat{\mathcal{I}}_g^{(1)}$ is determined by the $K_1$ largest values in $\mathbf{y}_g$. Similarly, $\hat{\mathcal{I}}_g^{(2)}$ is equivalent to a set of $K_2$ smallest entries in $\mathbf{y}_g$, except $y_{g, j}$ for $j \in \hat{\mathcal{I}}_g^{(1)}$.

This OS decoder is particularly useful for the fast decoding because the sorting algorithm requiring a complexity of $\mathcal{O}(\log(K_1+K_2)  M)$ is sufficient to decode ${\bf \hat x}_g$ from ${\bf y}_g$. In our simulations, we shall use this OS decoder for implementation efficiency.

\section{Performance Analysis}
In this section, we derive an exact analytical expression for the BLERs of single-layered BOSS codes with the proposed two-stage MAP decoding. Our analysis provides an insight into how the BLER changes according to blocklength $M$, the number of unitary matrices $G$, and $\frac{ E_{\rm b} }{ N_0 }$. The following theorem is the main result of this section.

\textbf{Theorem 1. } Let $\mathcal{E}_1$ and $\mathcal{E}_2$ be the error events for the first and second stage decoding. The BLER of a single-layered BOSS code with $\mathcal{A} = \{ 1 \}$ and rate $R = \frac{\lfloor \log_2 (M) \rfloor + \lfloor \log_2 (G) \rfloor}{M}$ is given by
\begin{align*}
    P_{\sf BLER} (M,G,\sigma_v^2) =  \mathbb{P} ( \mathcal{E}_1 ) + \mathbb{P} (\mathcal{E}_2 | \mathcal{E}_1^c ) \mathbb{P} (\mathcal{E}_1^c ),
    \numberthis{\label{BLER_analytic_expression}}
\end{align*} where 
\begin{align}
    \mathbb{P} ( \mathcal{E}_1 ) =  1 - \frac{(M - 1)}{\sqrt{2\pi \sigma_v^2}} \int_{-\infty}^\infty Q \left( \frac{y - 1}{\sigma_v} \right) \left[ 1 - Q \left( \frac{y}{\sigma_v} \right) \right]^{M - 2} e^{-\frac{y^2}{2\sigma_v^2}} {\rm d} y , \label{BLER_analytic_stage1}
\end{align}
and
\begin{align}
   \mathbb{P} (\mathcal{E}_2 | \mathcal{E}_1^c ) =  1-\left[\frac{ \Gamma \left( \frac{M}{2} \right)}{\sqrt{\pi} \Gamma \left( \frac{M - 1}{2} \right)} \int_{-1}^1  Q \left( \frac{ w-1}{\sqrt{2 \sigma_v^2}} \right)  (1 - w^2)^\frac{M - 3}{2} {\rm d}w\right]^{M(G-1)}. \label{BLER_analytic_stage2}
\end{align}

\textit{Proof. } See Appendix A.

Although the BLER expression in Theorem 1 is integral form, it illuminates how the decoding performance of BOSS codes is affected by code parameters. Note that the noise power $\sigma_v^2$ is related to $\frac{ E_{\rm b} }{ N_0 }$ and code rate $R=\frac{\lfloor \log_2 (M) \rfloor + \lfloor \log_2 (G) \rfloor}{M}$ as
\begin{equation}
    \sigma_v^2 = \frac{1}{M} \frac{1}{ 2 R \frac{ E_{\rm b} }{ N_0 } } = \frac{1}{2(\lfloor \log_2 (M) \rfloor + \lfloor \log_2 (G) \rfloor)\frac{ E_{\rm b} }{ N_0 }}. \label{eq:noise} 
\end{equation}

First, for fixed $\frac{ E_{\rm b} }{ N_0 }$ and $G$, it is evident that both $\mathbb{P} ( \mathcal{E}_1 ) $ and $\mathbb{P} ( \mathcal{E}_2 )$ decrease as  blocklength $M$ increases because the effective noise power $\sigma_v^2$ is inversely proportional to $M$, and $\mathbb{P} ( \mathcal{E}_1 ) $ and $\mathbb{P} ( \mathcal{E}_2 )$ are decreasing functions with respect to $M$. This confirms our intuition that the BOSS code with a large blocklength can improve the BLER performance. Second, for fixed $M$, increasing the number of unitary blocks $G$ impacts the BLERs differently depending on the noise power $\sigma_v^2$. When $\frac{ E_{\rm b} }{ N_0 }$ is sufficiently high, the second stage decoding error can vanish, i.e., $\lim_{\sigma_v^2\rightarrow 0}\mathbb{P} (\mathcal{E}_2 | \mathcal{E}_1^c ) \rightarrow 0$. This is somewhat surprising because increasing the code rate by using more $G$ unitary blocks does not occur any second stage decoding error, provided that $\frac{ E_{\rm b} }{ N_0 }$ is high enough. In this case, the resultant BLER is bounded by the first stage decoding error. However, when $\frac{ E_{\rm b} }{ N_0 }$ is not sufficiently high, increasing $G$ deteriorates the second stage decoding performance. Lastly, for fixed $G$, and $M$, it is  apparent that the decoding error approaches zero as increasing $\frac{ E_{\rm b} }{ N_0 }$. This also confirms that our BOSS code with the two-stage decoder does not have error-floor effects, which is particularly useful for URLLC applications such as telesurgery and augmented reality wherein accurate, haptic feedback is required, so a target BLER is extremely low (e.g., $10^{-6}$).

\section{CA-BOSS Codes}
% shall elaborate more on benefits of CRC concatenation. show a contrast btw list decoding with and without CRC!!
% 1. decrease the # of codeword candidates to be computed
% 2. filter out error events that invalid combinations happen to be closer to the measurement

In this section, we present a serially concatenated scheme of BOSS codes with short CRC as the outer code. This new family of BOSS codes, namely CA-BOSS codes, further improves the decoding performance with a simple list decoder, while increasing the decoding complexity very marginally.

\subsection{Concatenated Encoding}
\begin{figure*}[h]
    \centering
    \includegraphics[width = 0.7\textwidth]{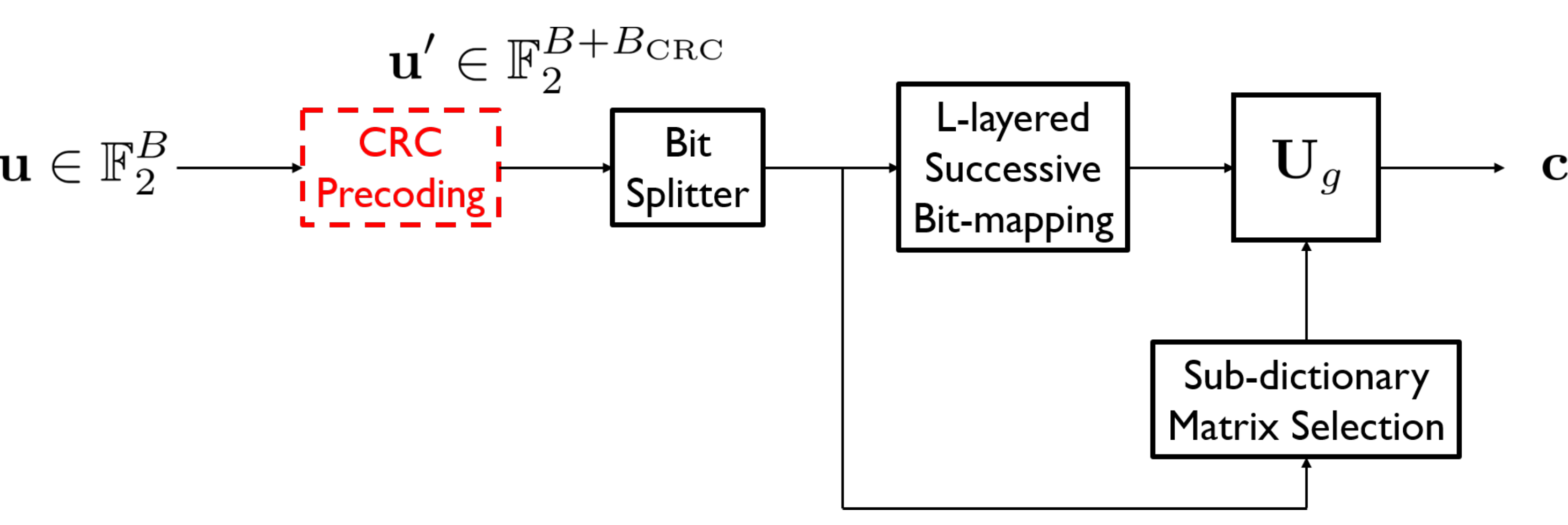}
    \caption{Concatenated encoding of CA-BOSS codes.}
    \label{Fig4:CA_BOSS_Encoder}
\end{figure*}

In the proposed MAP-based decoding algorithm, wrong decisions made in the early phases of the first stage cannot be fixed and result in incorrect candidate sets for the remaining layers. We adopt CRC precoding to help the decoder determine the non-zero support. $B_\text{CRC}$ CRC bits are added to the data block $\mathbf{u} \in \mathbb{F}_2^B$. The appended block $\mathbf{u}' \in \mathbb{F}_2^{B + B_\text{CRC}}$ is then fed into the original BOSS encoder, and its fractions are sequentially mapped into a dictionary block index and non-zero elements. The CA-BOSS encoder is illustrated in Fig. \ref{Fig4:CA_BOSS_Encoder} with a block newly added to the original BOSS encoder colored red.

% In the proposed MAP-based decoding algorithm, wrong decisions made at early phases of the first stage cannot be fixed and also result in incorrect candidate sets for the remaining layers. Moreover, from simulation runs, we have observed that erroneous detection mostly occurs at the first stage, in particular at low SNR. That is, once the sparse message vector is correctly estimated, the decoder readily identifies a true block index. This observation conforms with our intuitive understanding of BOSS codes. Each sub-dictionary matrix $\mathbf{U}_g$ has their own codeword ensemble, and even if the decoder finds the true support, i.e., $\hat{\mathcal{I}}_g = \mathcal{I}$, for every hypothesis, codeword estimates $\hat{\mathbf{c}}_g$ except the true block index can be thought to reside in distinct code-spaces, thereby being greatly distinguishable from $\mathbf{y}$.  

\subsection{List Decoder}
\begin{figure*}[h]
    \centering
    \includegraphics[width=0.7\textwidth]{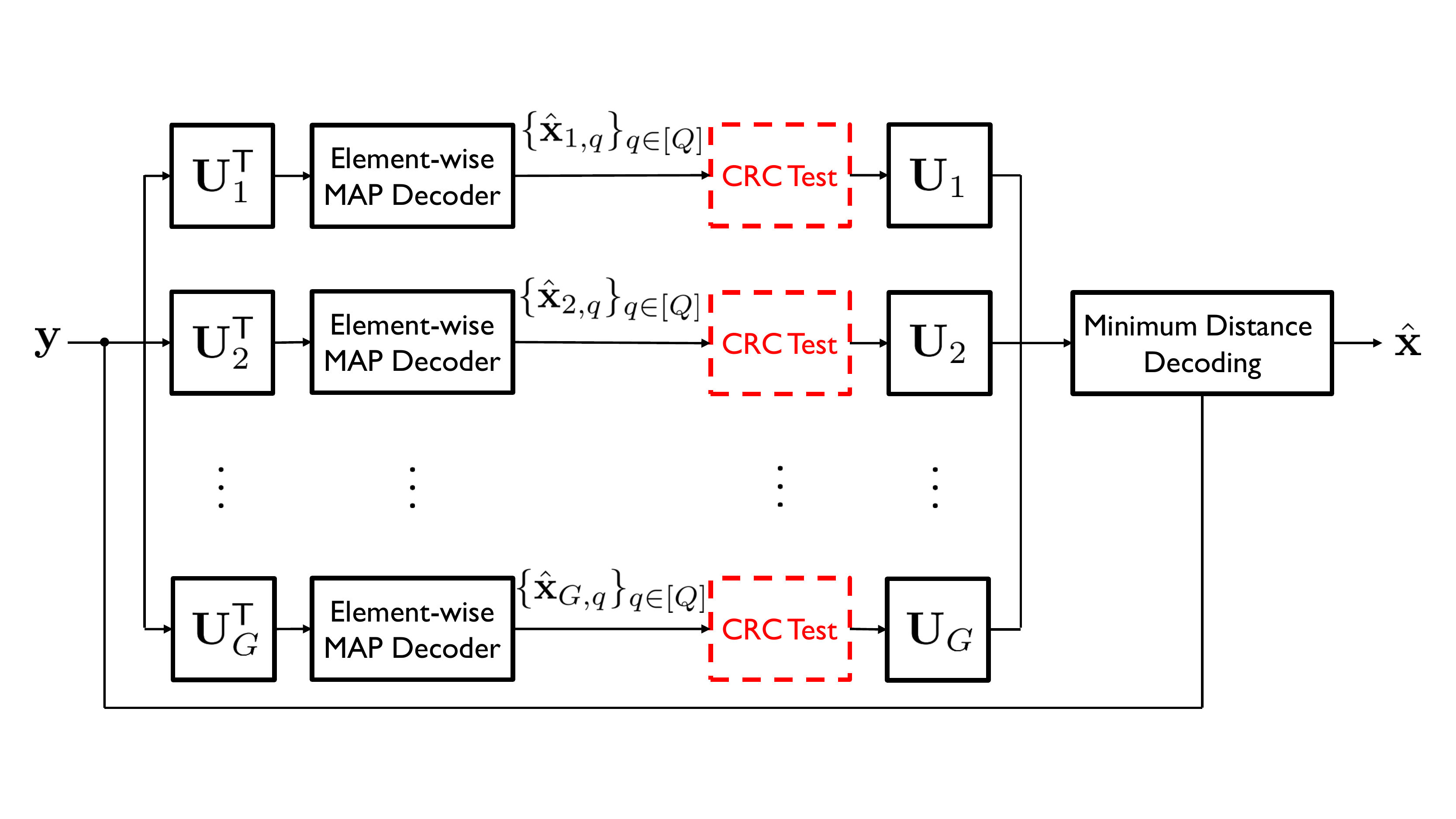}\vspace{-0.5cm}
    \caption{The MAP-List decoder design of CA-BOSS codes.}
    \label{Fig5:CA_BOSS_ListDecoder}
\end{figure*}

We introduce a list decoding algorithm based on the two-stage MAP decoder to take advantage of CRC error-detection. Under each $\mathscr{H}_g$, the decoder finds $Q$ sub-message vector estimates, i.e., $\left\{ \hat{\mathbf{x}}_{g, 1} , \hat{\mathbf{x}}_{g, 2} , \dots , \hat{\mathbf{x}}_{g, Q} \right\}$. The list can be generated in various ways. For example, when a two-layered BOSS code with $K_1 = K_2 = 1$ is evaluated, the decoder can select the two most likely indices at each layer and obtain a list of $Q = 4$, using their combinations. The decoder inverse-maps each provisional support $\{ \hat{\mathcal{I}}_{g, q} \}_{q \in [Q]}$ and $g$ to a bit sequence and performs a CRC test to identify a valid combination of the block and non-zero indices. Only surviving message vector estimates are transferred to the next decoding stage and re-encoded into tentative codewords. Finally, the decoder determines the block index and most probable valid support by choosing the codeword candidate closest to $\mathbf{y}$. The proposed MAP-List decoding is described in Fig. \ref{Fig5:CA_BOSS_ListDecoder}.

\section{Numerical Results}
In this section, we provide simulation results to validate the error-correcting performance of proposed BOSS codes under the two-stage MAP decoding algorithm. We then investigate how CRC concatenation affects the performance of BOSS codes.

\subsection{BLER Comparison in the AWGN Channel}
We adopt BLER versus $E_{\rm b}/N_{\rm 0}$ as an assessment of the two-stage MAP decoder in the AWGN channel to take into consideration low code rates and non-standard modulation techniques of BOSS codes. 

\begin{figure*}[h]
    \centering
    \includegraphics[width=0.6\textwidth]{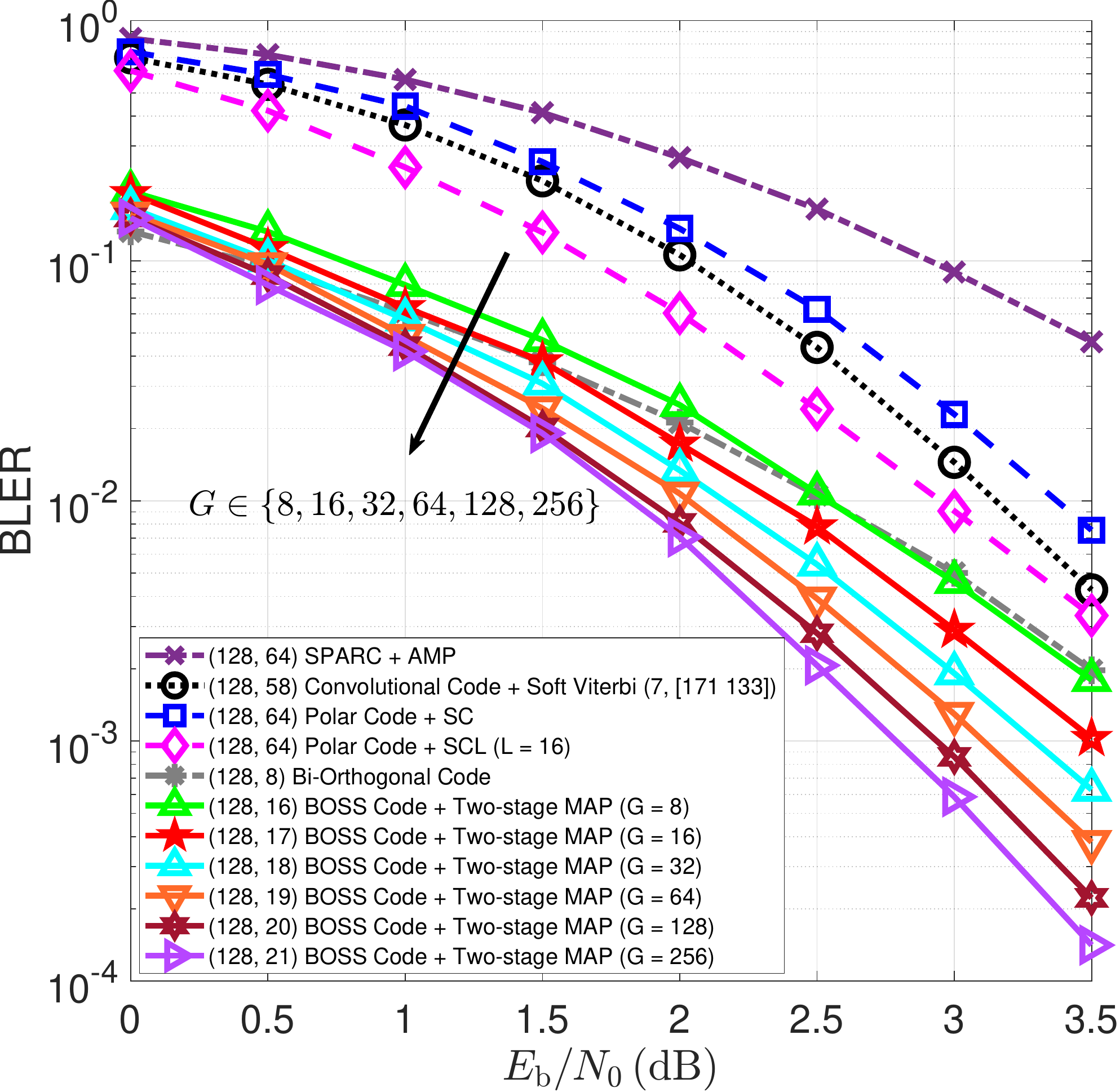}
    \caption{BLER comparison with existing codes.}
    \label{Fig6:BOSS_AWGN_BLER}
\end{figure*}

\textbf{BOSS codes: } We simulated two-layered BOSS codes with blocklength $M = 128$; symmetric sparsity $K_1 = K_2 = 1$; singleton PAM alphabets $\mathcal{A}_1 = \{ 1 \}$ and $\mathcal{A}_2 = \{ -1 \}$; and various block sizes. For $G = 16$, the code rate is $\frac{17}{128} \approx 0.13$.

\textbf{Polar codes: } We considered a $1/2$ rate polar code of length $128$, using the binary phase-shift keying (BPSK) modulation. The code construction is based on the $2 \times 2$ Ar{\i}kan kernel, and the 3GPP 5G NR frozen set, i.e., pre-computed SNR-independent channel reliability order, \cite{3GPP} is adopted. SC and SCL decoding algorithm with search width $L = 16$ were employed.

\textbf{Convolutional codes: } A convolutional code of length $128$ with BPSK modulation was evaluated under the soft Viterbi algorithm. The generator polynomials are $\textsl{g} = [ 171 \, 133]$ in octal notation, so the constraint length is $7$. Taking account of rate loss entailed by termination bits, the effective code rate is $\frac{58}{128} \approx 0.45$. 

\textbf{SPARC codes: } We used a fat random Gaussian dictionary matrix $\mathbf{A}$ consisted of $8$ sections to generate a SPARC of length $128$. Each section contains $256$ columns, i.e., $\mathbf{A} \in \mathbb{R}^{128 \times 2048}$, so the code rate is $\frac{8 \log_2 256}{128} = \frac{1}{2}$. 

\textbf{Bi-orthogonal codes: } We considered a bi-orthogonal code of length $128$ with a dictionary matrix $\mathbf{A} = [ \mathbf{H}_{2^7}, -\mathbf{H}_{2^7} ],$ where $\mathbf{H}_{2^k}$ is a Hadamard matrix of order $2^k$. Each codeword conveys $\log_2 (2 \times 128) = 8$ bits in $128$ channel uses.

Fig. \ref{Fig6:BOSS_AWGN_BLER} shows that BOSS codes outperform polar codes under SCL decoding by a considerable margin: the proposed decoder results in approximately $0.4$ coding gains at BLER of $10^{-4}$ even for $G = 8$. The results in Fig. \ref{Fig6:BOSS_AWGN_BLER} also illustrate a prominent feature of BOSS codes: enlarging the dictionary matrix improves decoding performance at the cost of computational complexity. This trend can be ascribed to an increase in the code rate caused by concatenating the dictionary matrix with more blocks, whereas the average transmit energy $E_s$ is preserved. Therefore, the block number parameter $G$ can be appropriately tuned to achieve a desired trade-off between performance and complexity. Surprisingly, no performance saturation was observed when increasing $G$ up to 256. A bi-orthogonal code shows comparable performance to that of our BOSS codes with a few blocks. Therefore, it can be thought as a special case of BOSS codes when $G=2$ and  $K=1$.

\subsection{Decoding Performance of CA-BOSS codes}
\begin{figure*}[h]
    \centering
    \includegraphics[width=0.6\textwidth]{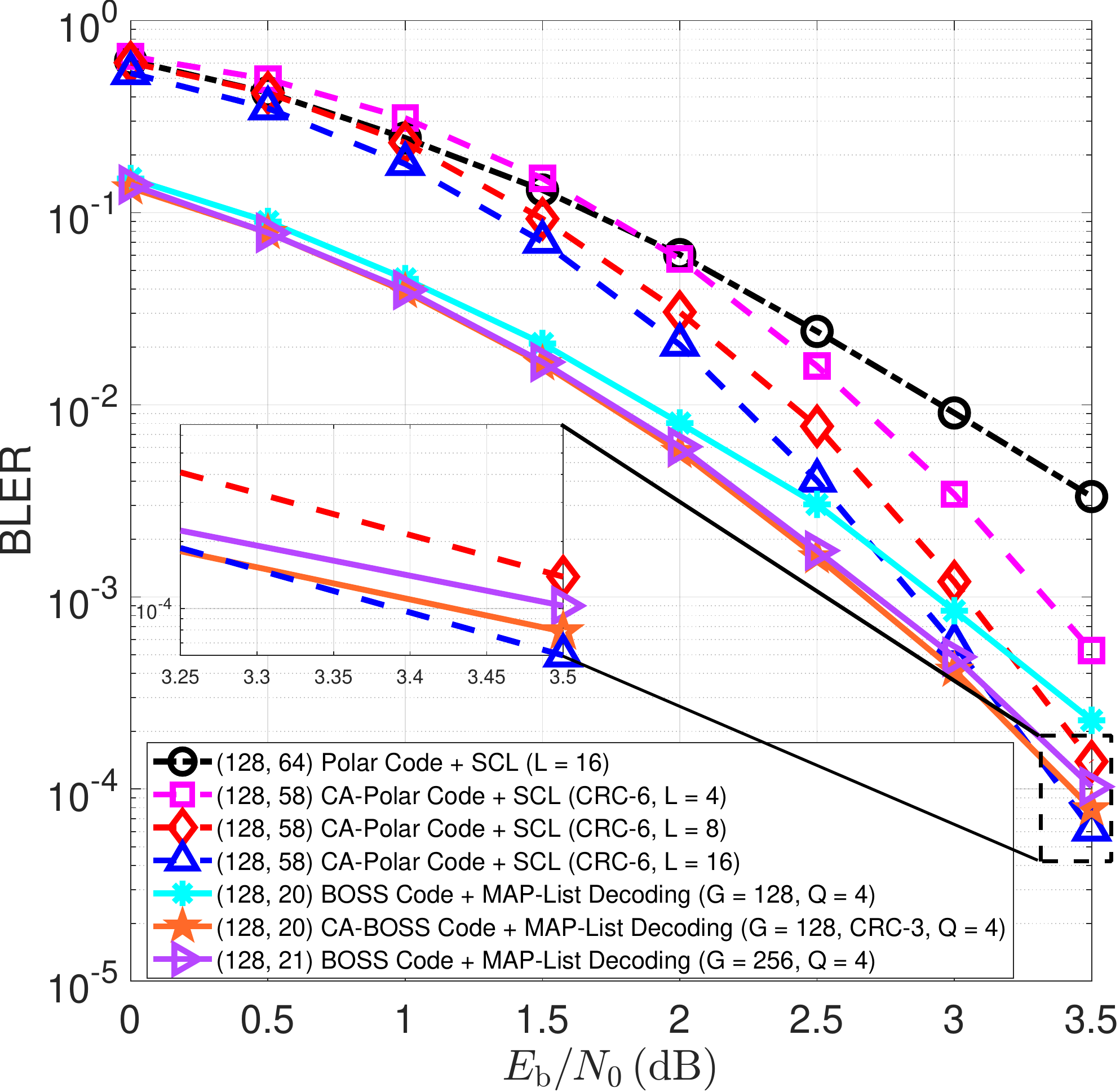}
    \caption{BLER performance comparison between CA-BOSS and CA-polar codes.}
    \label{Fig7:CA_BOSS_BLER_EbN0}
\end{figure*}

Fig. \ref{Fig7:CA_BOSS_BLER_EbN0} compares CA-BOSS codes to the state-of-the-art CA-polar codes under SCL decoding with different search width. It can be seen that CA-BOSS codes outperform ordinary BOSS codes of the same rate by a meaningful margin. Moreover, at high SNR, the CA-BOSS code with $G = 128$ even shows superior decoding performance to the BOSS code with $G = 256$. We observed from the AWGN simulation result that the performance-complexity trade-off of BOSS coding is controlled by $G$. This favorable trait, however, does not lend itself to CA-BOSS codes, since the effective code rate remains the same. If the list decoding is adopted without CRC concatenation, the decoder at the second stage will generate $Q$ codeword candidates per hypothesis, and incorrect estimates happen to be closer to the observation. The CRC test is capable of ruling out error events caused by invalid support estimates under the true hypothesis. Consequetly, CA-BOSS codes enjoy two gains in reducing decoding complexity by the codeword pruning and enhancing the decoding performance by error detection.

In the low SNR regime, CA-BOSS codes show excellent decoding performance; however, as the SNR increases, BLER of CA-polar codes drops with a precipitous slope, beating CA-BOSS codes.

\begin{figure*}[h]
    \centering
    \includegraphics[width=0.6\textwidth]{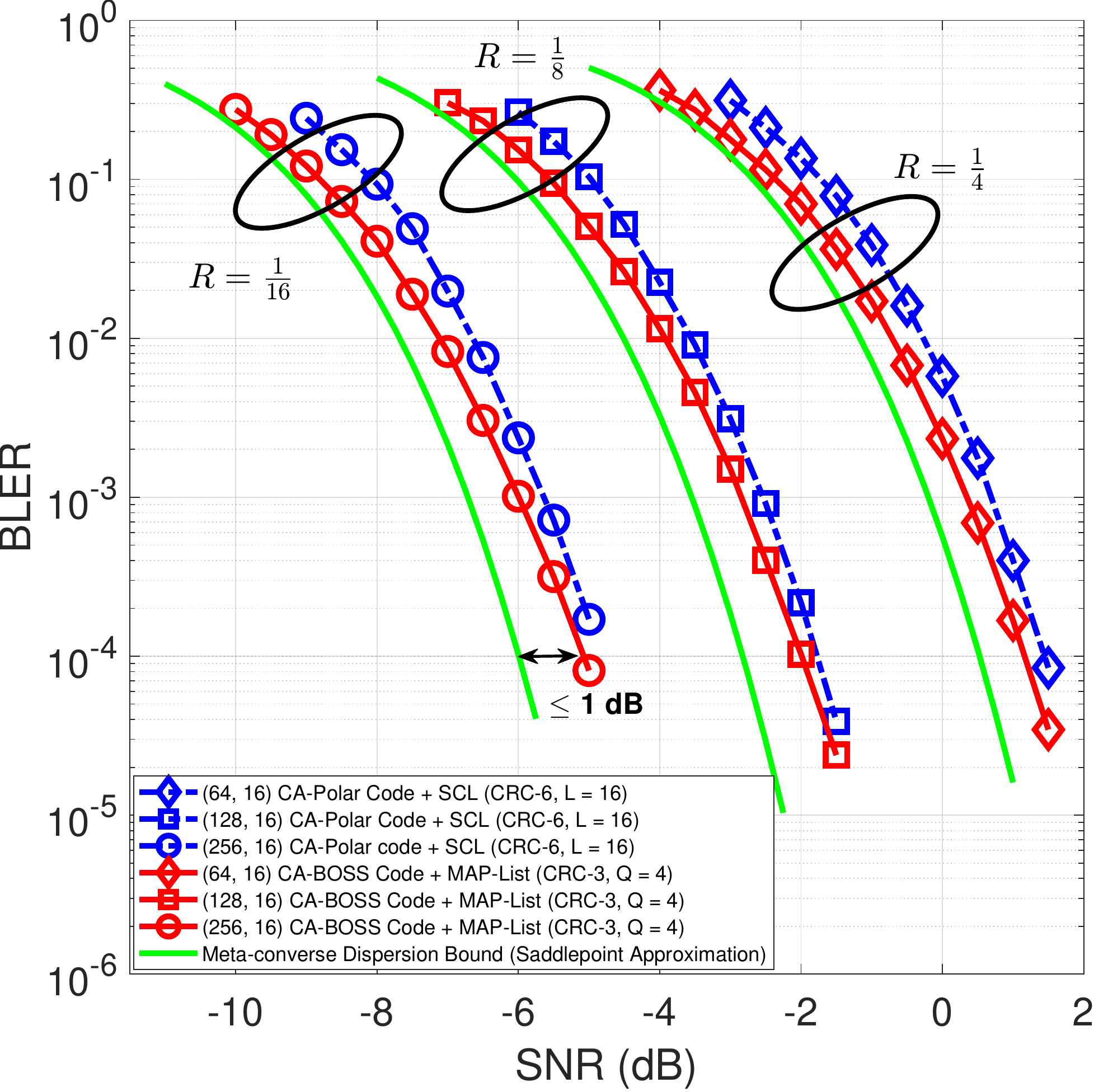}
    \caption{BLER performance comparison for BOSS and polar codes when $R\in \{1/4,1/8,1/16\}$.}
    \label{Fig8:CA_BOSS_BLER_SNR}
\end{figure*}

Fig. \ref{Fig8:CA_BOSS_BLER_SNR} compares CA-BOSS codes with CA-polar codes of the same rate. It turns out that CA-BOSS codes outperform the polar counterparts at every blocklength under consideration, even when attached with the shorter, i.e., weaker, CRC outer code. This remarkable result attests to the error-correction capability of our codes. When transmitting the same information block size, adding CRC bits causes the polar encoder to use additional non-extremal (having mediocre reliability) virtual bit-channels. As a result, the receiver is more prone to errors in the early phase of sequential decoding. This fact, however, is not a concern of CA-BOSS coding.

Fig. \ref{Fig8:CA_BOSS_BLER_SNR} also plots the saddlepoint approximation \cite{metaConverseBound_saddlePtApprox} to the meta-converse bound \cite{Polyanskiy_finiteCapacity}: 
\begin{align}
    \eta(n, M) \overset{\triangle}{=} \underset{P^n}{\min} \, \underset{Q^n}{\max} \{ \alpha_{\frac{1}{M}} ( P^n \times W^n, P^n \times Q^n ) \},
\end{align} where $M$ is the number of symbols transmitted over a length-$n$ $W^n (\cdot | \cdot )$ channel, and $\alpha_\beta (P, Q)$ is the smallest type-I error probability across all tests between an input distribution $P$ and auxiliary output distribution $Q$, with a type-II error probability of at most $\beta$. It can be seen that our CA-BOSS codes perform within one dB away from the finite-length channel capacity bounds in all SNRs and  rates.

\subsection{Validation of BLER Analysis}

\begin{figure*}[h]
    \centering
    \begin{subfigure}[t]{0.49\textwidth}
        \centering
        \includegraphics[width=1\textwidth]{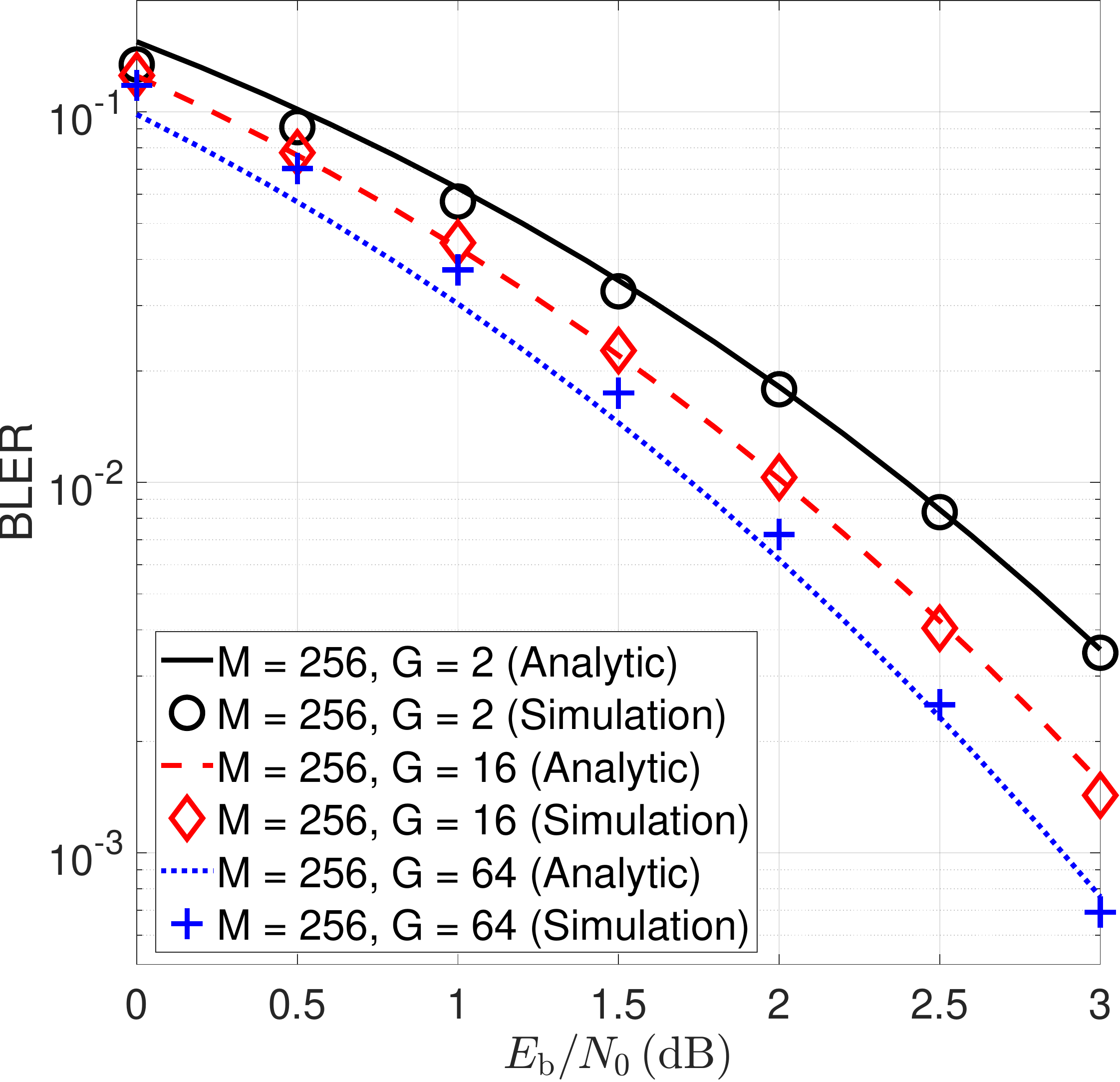}
        \caption{}
        \label{Fig9a:BLER_Analysis_M256}
    \end{subfigure}%
    \hfill
    \begin{subfigure}[t]{0.49\textwidth}
        \centering
        \includegraphics[width=1\textwidth]{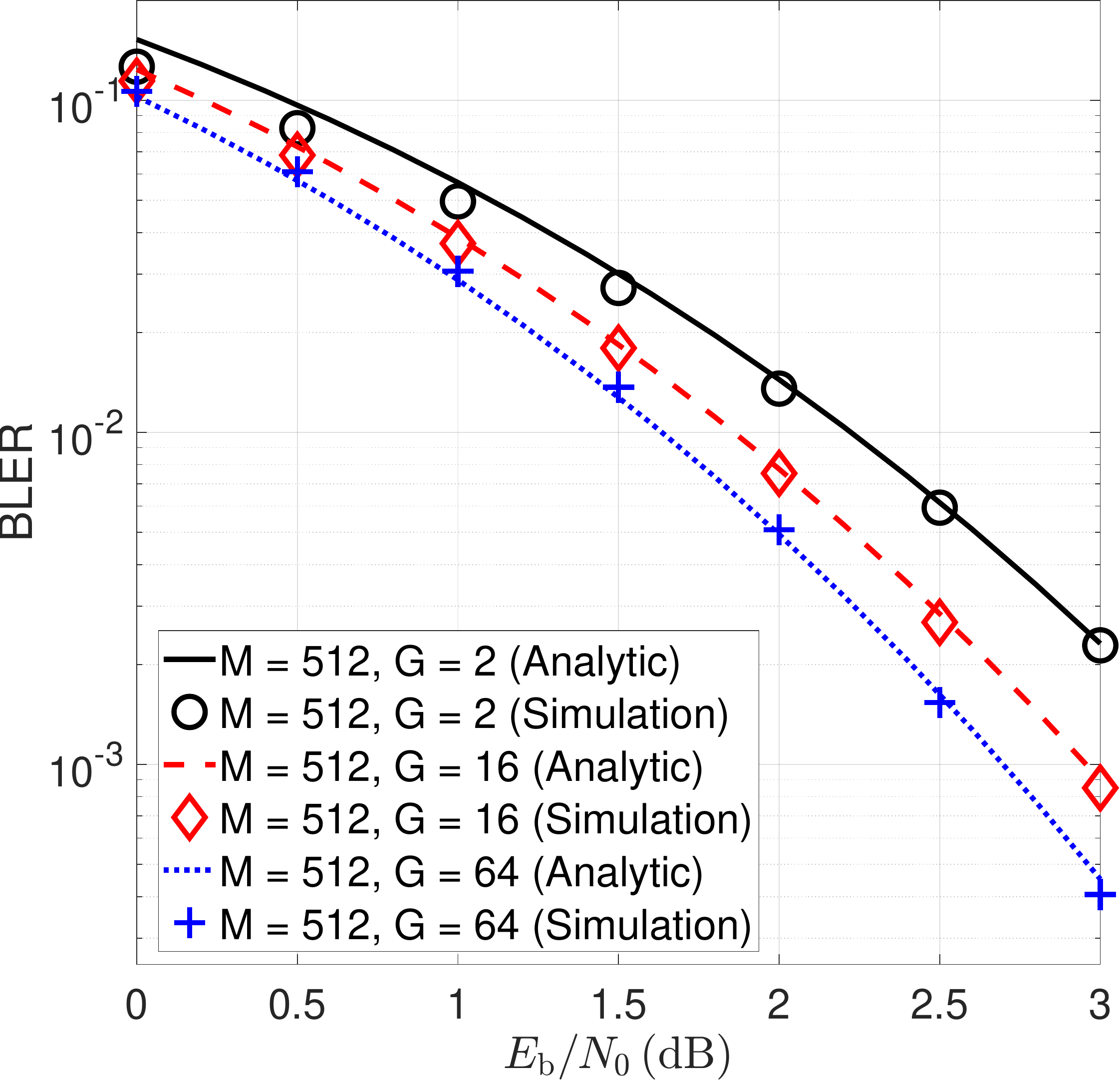}
        \caption{}
        \label{Fig9b:BLER_Analysis_M512}
    \end{subfigure}
    \caption{BLER comparison between analytical and simulation results for $M\in \{256,512\}$ and $G\in \{2,16,64\}$.}
    \label{Fig9:BLER_Analysis}
\end{figure*}

% \begin{figure*}[t]
%     \centering
%     \includegraphics[width=0.6\textwidth]{BLER_Analysis.pdf}
%     \caption{Comparison between empirical data and analytical BLERs of $[M, G, K] = [128, 2, 1]$ BOSS code}
%     \label{Fig9:BLER_Analysis}
% \end{figure*}

Fig. \ref{Fig9:BLER_Analysis} compares the analytical BLER derived in Theorem 1 with the simulations for $M=\{256,512\}$, $\mathcal{A} = \{ 1 \}$, and $G=\{2,16,64\}$. As can be seen, our analytical expression tightly matches with the simulation results for various code parameters, which confirms the exactness of our analysis. In the low $E_b/N_0$ regime, our analytical results show a small discrepancy with simulations. This gap arises from the numerical precision errors in computing $M(G-1)$ power in \eqref{BLER_analytic_stage2}, and it becomes pronounced as $M(G-1)$ increases. Nonetheless, this gap readily vanishes as $E_b/N_0$ increases. 

%Finding another analytical BLER expression that is resilient to the numerical precision errors is an Using extreme value theory, 

% \textbf{Remark (Design principle): } Many modern codes fully exploit the encoding structure to design an effective decoding algorithm. Convolutional coding scheme creates a natural trellis diagram that facilitates the implementation of the Viterbi decoder. Belief propagation decoding of LDPC codes leans on the sparse nature. 

\section{Conclusion}

From the point of view of coded modulation techniques, this work has provided a new type of joint coded modulation method called BOSS codes for URLLC. In particular, we have presented a novel successive encoding technique to generate zero-RIP codewords that are a sparse linear combination of orthogonal columns of a dictionary matrix. A very fast yet noise-tolerant MAP-based decoding algorithm has been proposed for BOSS codes. This two-stage MAP decoder exploits the orthogonality and performs element-wise MAP decoding to identify non-zero coefficients in a parallel structure, thereby achieving near-ML decoding performance with low decoding latency. We then provide the exact BLER expression for single-layered BOSS codes. This analytical formula is useful in understanding on how crucial code parameters affect the performance of BOSS codes. We have further improved the performance of BOSS codes through concatenation with a CRC outer code and shown that CA-BOSS codes achieve the meta-converse bound within one dB in the low code-rate regime.  

In recent follow-up work, a robust decoding algorithm for BOSS codes was studied in multi-path fading environments, and the real-time demonstration using NI-PXI according to IEEE 802.11 standards was shown to considerably outperform CA-polar codes with SCL decoding when a codeword transmission experiences multiple fading states \cite{BOSS_Fading1, BOSS_Fading2}. A promising direction for future work is  to develop a joint equalizer and decoder that is robust for the multi-fading states. It is also an interesting direction to expand BOSS codes by incorporating phase shift keying modulation in complex AWGN channels to deliver more information bits in an energy-efficient manner.

%Multi-path fading and intersymbol interference are expected to destroy the intrinsic orthogonal property of BOSS codes, thereby rendering the aforementioned two-stage MAP decoder invalid. In particular, rate loss and pilot overhead cannot be neglected in short-packet transmissions, so non-coherent detection will pose an interesting research challenge.

\section*{Appendix}
\subsection{Proof of Theorem 1}
Before providing the proof, we first introduce two lemmas, which are necessary for our proof. The first lemma provides probability distribution of the maximum and minimum of $M$ IID random variables.

\textbf{Lemma 1. } \textit{Let $\{ X_m \}_{m=1}^M$ be a sequence of $M$ IID random variables, each with probability density function (PDF) $f_X (x)$ and cumulative density function (CDF) $F_X (x)$. We also denote the maximum and minimum of the sequence by} $X_\text{max} = \underset{1 \leq m \leq M}{\max} \, X_m$ \textit{and} $X_\text{min} = \underset{1 \leq m \leq M}{\min} \, X_m$ \textit{, respectively. Then, the PDF's of} $X_\text{max}$ \textit{and} $X_\text{min}$ \textit{are}
\begin{align}
    f_{X_\text{max}} (x) &= M f_X (x) F_X (x)^{M - 1} \label{max_distribution} \quad \text{and} \\
    f_{X_\text{min}} (x) &= M f_X (x) [1 - F_X (x)]^{M - 1}. \label{min_distribution}
\end{align}

\textit{Proof. } Due to independence, the CDF of $X_\text{max}$ can be readily found:
\begin{equation}
    F_{X_\text{max}} (x) = \prod_{m=1}^M \mathbb{P} (X_m \leq x) = F_X (x)^M . \label{max_CDF}
\end{equation} By differentiating $F_{X_\text{max}} (x)$ with respect to $x$, we obtain $f_{X_{\text{max}}} (x)$. Similarly, $F_{X_\text{min}(x)} = 1 - \left[ 1 - F_X (x) \right]^M$, and the rest of the proof is straightforward.

The second lemma provides statistics of an inner product of two IID random unit vectors in an $(M - 1)$-dimensional sphere.

\textbf{Lemma 2. } \textit{For two random unit-norm vectors, $\mathbf{u} \in \mathbb{R}^M$ and $\mathbf{v} \in \mathbb{R}^M$, that are uniformly distributed on a sphere $\mathcal{S}^{M - 1}$, their inner product $W = \langle \mathbf{u} , \mathbf{v} \rangle$ has a PDF given by}
\begin{equation}
    f_W (w) = \frac{ \Gamma \left( \frac{M}{2} \right)}{\sqrt{\pi} \Gamma \left( \frac{M - 1}{2} \right)} (1 - w^2)^{\frac{M - 3}{2}} .  \label{innerproduct_pdf} 
\end{equation}

\textit{Proof. } If $\mathbf{X} = \{ X_1, X_2, \dots X_M \}$ follows the standard multivariate normal distribution, i.e., $X_i \sim \mathcal{N}(0, 1)$, then $\frac{\mathbf{X}}{\| \mathbf{X} \|_2}$ is uniformly distributed on the unit sphere. We assume that $\mathbf{v} = [1, 0, \dots, 0]$ thanks to spherical symmetry, and then the distribution of $W$ is identical to that of 
\begin{equation}
    \frac{X_1}{\sqrt{X_1^2 + X_2^2 + \cdots + X_M^2}}. \label{ratio_distribution}
\end{equation} It is well known that the square of ratio distribution in \eqref{ratio_distribution} follows $\text{Beta} ( \frac{1}{2} , \frac{M - 1}{2} )$. The rest of the proof is trivial.

Now, we are ready to prove Theorem 1. We denote by $\mathcal{E}_1$ and $\mathcal{E}_2$ an error event at stage 1 and 2 of the proposed two-stage MAP decoder, respectively. Using these error events, the probability of decoding error is given by
\begin{equation}
    \mathbb{P}(\mathcal{E}) = \mathbb{P}(\mathcal{E}_1) + \mathbb{P}(\mathcal{E}_2 | \mathcal{E}_1^c ) \mathbb{P}(\mathcal{E}_1^c ) . \label{decoding_error_probability}
\end{equation} We consider a randomly constructed dictionary matrix $\mathbf{A} = \left[ \mathbf{U}_1 , \mathbf{U}_2 , \dots , \mathbf{U}_G  \right]$, i.e., $\mathbf{U}_i$ and $\mathbf{U}_j$ are independent for $i \ne j \in [G]$. Without loss of generality, we assume that $K_1 = 1$ and $\mathbf{U}_1$ has participated in encoding, so $\mathbf{A} \mathbf{x} = \mathbf{U}_1 \mathbf{x}_1$, where $\|\mathbf{x}_1\|_0 = 1$. For the AWGN channel, the received symbol in the $m$th channel use is given by
\begin{equation}
    Y_m = \begin{cases}
    1 + V_m & \text{for } m \in \mathcal{I} \\
    V_m & \text{for } m \notin \mathcal{I} ,
    \end{cases} \label{AWGN_receivedSymbol}
\end{equation} where $V_m$ is a zero-mean Gaussian noise with variance $\sigma_v^2$. As explained in Remark 4, a simple OS decoder is optimal in this case. Then, the error event at stage is $\{ \hat{\mathbf{x}}_1 \ne \mathbf{x}_1 \}$, and its probability can be computed as 
\begin{align*}
    \mathbb{P}( \mathcal{E}_1 ) &= 1 - \mathbb{P} ( \mathcal{E}_1^c ) \\ 
    &= 1 - \mathbb{P} (Y_\mathcal{I}^\text{min} > Y_{\mathcal{I}^c}^\text{max} ) , \numberthis{\label{stage1_error_prob_decomposition}}
\end{align*} where $Y_\mathcal{I}^\text{min} \overset{\triangle}{=} \underset{m \in \mathcal{I}}{\min} \, Y_m$ and $Y_{\mathcal{I}^c}^\text{max} \overset{\triangle}{=} \underset{m \in \mathcal{I}^c}{\max} \, Y_m$. The conditional expectation theorem gives
\begin{align*}
    \mathbb{P}(Y_\mathcal{I}^\text{min} > Y_{\mathcal{I}^c}^\text{max}) &= \mathbb{E}_{Y_{\mathcal{I}^c}^\text{max}} \left\{ \mathbb{P} ( Y_\mathcal{I}^\text{min} > y ) | Y_{\mathcal{I}^c}^\text{max} = y \right\} \\
    &= \int_{-\infty}^\infty \mathbb{P} ( Y_\mathcal{I}^\text{min} > y ) f_{ Y_{\mathcal{I}^c}^\text{max} } (y) {\rm d} y \\
    &= \int_{-\infty}^\infty \mathbb{P} ( Y_\mathcal{I}^\text{min} > y ) (M - 1) f_{Y_{\mathcal{I}^c}} (y) F_{Y_{\mathcal{I}^c}} (y)^{M - 2} {\rm d} y \\
    &= \frac{(M - 1)}{\sqrt{2\pi \sigma_v^2}} \int_{-\infty}^\infty [1 - F_{Y_\mathcal{I}}(y)] F_{Y_{\mathcal{I}^c}} (y)^{M - 2} e^{-\frac{y^2}{2\sigma_v^2}} {\rm d}y. \numberthis{\label{stage1_OS_decoding_error}}
\end{align*} Note that $Y_m$ is distributed as $\mathcal{N}(1, \sigma_v^2)$ for $m \in \mathcal{I}$ and $\mathcal{N}(0, \sigma_v^2)$ for $m \in \mathcal{I}^c$. Plugging $F_{Y_\mathbb{I}}(y) = 1 - Q \left( \frac{y - 1}{\sigma_v} \right)$ and $F_{Y_{\mathcal{I}^c}} (y) = 1 - Q \left( \frac{y}{\sigma_v} \right)$ into  \eqref{stage1_OS_decoding_error}, we arrive at the expression in \eqref{BLER_analytic_stage1}.
% \begin{equation}
%     \mathbb{P} (\mathcal{E}_1 ) = 1 - \frac{(M - 1)}{\sqrt{2 \pi \sigma_v^2}} \int_{-\infty}^\infty Q \left( \frac{y - 1}{\sigma_v} \right) \left[ 1 - Q \left( \frac{y}{\sigma_v} \right) \right]^{M - 2} e^{-\frac{y^2}{2\sigma_v^2}} {\rm d} y. \label{stage1_error_prob}
% \end{equation}

Now, we shift our focus to the second stage. Under the condition that the decoder has correctly estimated a sparse sub-message vector under $\mathscr{H}_1$, i.e., $\hat{\mathbf{x}}_1 = \mathbf{x}_1$, an error occurs when at least one codeword estimate $\mathbf{U}_g \hat{\mathbf{x}}_g$ for $g \in [G]\backslash \{1\}$ is closer to $\mathbf{y}$ in terms of Euclidean distance compared to $\mathbf{U}_1 \hat{\mathbf{x}}_1$. The conditional error probability at stage 2 can be expressed as follows:
\begin{align*}
    \mathbb{P}(\mathcal{E}_2 | \mathcal{E}_1^c ) &= 1 - \mathbb{P} ( \hat{g} = 1 | \hat{\mathbf{x}}_1 = \mathbf{x}_1 ) \\
   % &= 1 - \mathbb{P} \left( \cap_{g \in [G]\backslash\{1\}} \{ \| \mathbf{y} - \mathbf{U}_1 \hat{\mathbf{x}}_1 \|_2 < \| \mathbf{y} - \mathbf{U}_g \hat{\mathbf{x}}_g \|_2 \} | \hat{\mathbf{x}}_1 = \mathbf{x}_1 \right) \\
    &= 1 - \mathbb{P} \left( \cap_{g \in [G] \backslash \{1 \}} \{ \mathbf{y}^\mathsf{T} \mathbf{U}_1 \mathbf{x}_1 > \mathbf{y}^\mathsf{T} \mathbf{U}_g \hat{\mathbf{x}}_g \} \right).  \numberthis{\label{stage2_conditional_error_prob}}
\end{align*} We first consider the probability that $\mathbf{y}^\mathsf{T} \mathbf{U}_1 \mathbf{x}_1$ is larger than $\mathbf{y}^\mathsf{T} \mathbf{U}_g \hat{\mathbf{x}}_g$. Recall that the OS decoder estimates the support under $\mathscr{H}_g$ by finding the maximum element of $\mathbf{U}_g^\mathsf{T} \mathbf{y} = \mathbf{U}_g^\mathsf{T} ( \mathbf{U}_1 \mathbf{x}_1 + \mathbf{v} )$. We define the $m$th element of random vector $\mathbf{U}_g^\mathsf{T} \mathbf{y}$ as
\begin{equation}
    \tilde{Y}_{g,m} = \mathbf{u}_{g, m}^\mathsf{T} \mathbf{v} +  \mathbf{u}_{g, m}^\mathsf{T} \mathbf{U}_1 \mathbf{x}_1. \label{random_variable_gth_hypothesis}
\end{equation} The first term $\tilde{V}_{g,m} = \mathbf{u}_{g,m}^\mathsf{T} \mathbf{v}$ is a zero-mean Gaussian random variable with variance $\sigma_v^2$, while the second term $W_{g,m} = \mathbf{u}_{g,m}^\mathsf{T} \mathbf{U}_1 \mathbf{x}_1$ is distributed per Lemma 2. The decoder finds the maximum index:
\begin{equation}
    \hat{\mathcal{I}}_g = \underset{m \in [M]}{\arg \, \max} \, \mathbf{u}_{g,m}^\mathsf{T} \mathbf{U}_1 \mathbf{x}_1 + \mathbf{u}_{g,m}^\mathsf{T} \mathbf{v}. \label{support_estimate_gth_hypothesis}
\end{equation} The decoder generates $\hat{\mathbf{x}}_g$ as a one-hot vector with a non-zero value at $\hat{\mathcal{I}}_g$. Therefore, $\mathbf{y}^\mathsf{T} \mathbf{U}_g \hat{\mathbf{x}}_g$ is distributed as the maximum of $\tilde{Y}_{g,m}$, i.e., $\max \, \left\{ \tilde{Y}_{g, 1} , \dots , \tilde{Y}_{g, M} \right\}$, and \eqref{stage2_conditional_error_prob} can be re-written as 
\begin{align*}
    \mathbb{P} (\mathcal{E}_2 | \mathcal{E}_1^c ) &= 1 - \mathbb{P} \left( \bigcap_{g \in [G] \backslash \{1 \} } \left\{ \mathbf{y}^\mathsf{T} \mathbf{U}_1 \mathbf{x}_1 > \max \left\{ \tilde{Y}_{g, 1} , \dots , \tilde{Y}_{g, M} \right\} \right\} \right) \\
 %   &= 1 - \mathbb{P} \left( \bigcap_{\substack{g \in [G] \backslash \{1 \} \\ m \in [M]}} \left\{ \mathbf{y}^\mathsf{T} \mathbf{U}_1 \mathbf{x}_1 > \tilde{Y}_{g, m} \right\} \right) \label{stage2_conditional_error_prob_max_max(b)}\\
    &= 1 - \mathbb{P} \left( \bigcap_{\substack{g \in [G] \backslash \{1 \} \\ m \in [M]}} \left\{ \mathbf{v}^\mathsf{T} \mathbf{U}_1 \mathbf{x}_1 + \| \mathbf{U}_1 \mathbf{x}_1 \|_2^2 > \tilde{V}_{g,m} + W_{g,m} \right\} \right). \numberthis{\label{stage2_conditional_error_prob_max_max(c)}}
\end{align*}
Note that $\tilde{Y}_{g,m} = \tilde{V}_{g,m} + W_{g,m}$ is IID for $g \in [G] \backslash \{1 \}$. Denoting $\tilde{Z}_{g,m} = \mathbf{v}^\mathsf{T} \mathbf{U}_1 \mathbf{x}_1 - \tilde{V}_{g,m}$, a zero-mean Gaussian with variance $2\sigma_v^2$, we obtain the following expression: 
\begin{align*}
    \mathbb{P} (\mathcal{E}_2 | \mathcal{E}_1^c ) &= 1 - \prod_{\substack{g \in [G] \backslash \{ 1 \} \\ m \in [M]}} \mathbb{P} ( \tilde{Z}_{g,m} > W_{g,m} - 1 ) \\
    &= 1 - \prod_{\substack{g \in [G] \backslash \{ 1 \} \\ m \in [M]}} \mathbb{E}_{W_{g,m}} \left\{ \mathbb{P} (\tilde{Z}_{g,m} > w - 1 ) | W_{g,m} = w \right\} \\
    &= 1 - \left[ \frac{\Gamma \left( \frac{M}{2} \right)}{\sqrt{\pi} \Gamma \left( \frac{M - 1}{2} \right)} \int_{-1}^1 Q \left( \frac{w - 1}{\sqrt{2\sigma_v^2}} \right) (1 - w^2)^\frac{M - 3}{2} {\rm d} w \right]^{M(G - 1)}. \numberthis{\label{stage2_conditional_error_prob_final}}
\end{align*} Finally, by plugging \eqref{BLER_analytic_stage1} and \eqref{stage2_conditional_error_prob_final} into \eqref{decoding_error_probability}, we complete the proof.

\bibliographystyle{IEEEtran}
\bibliography{myRef}

\end{document}